\documentclass[eqsecnum]{elsart}
\usepackage{amssymb}
\usepackage{amsmath}
\usepackage[dvips]{graphicx}
\usepackage{dcolumn} 
\usepackage{bm, bbm} 
\usepackage{curves}
\usepackage{epic}
\usepackage{epsf, times}
\usepackage{subfigure}
\usepackage{psfrag}
\usepackage{natbib}
\usepackage{cmmib57}
\usepackage{euscript}
\newcommand{\onlinecite}{\cite}
\begin{document}

\begin{frontmatter}

\title{
Zero-temperature Kosterlitz-Thouless transition in a 
two-dimensional quantum system
      }

\author[addr1]{Claudio Castelnovo\corauthref{cor}\thanksref{CC and CC}}
\corauth[cor]{Corresponding author.} 
\ead{castel@buphy.bu.edu}
\author[addr1]{Claudio Chamon\thanksref{CC and CC}}
\ead{chamon@buphy.bu.edu}
\author[addr2]{Christopher Mudry}
\ead{christopher.mudry@psi.ch}
\author[addr3]{Pierre Pujol}
\ead{pierre.pujol@ens-lyon.fr}

\address[addr1]{Physics Department, Boston University, Boston, MA 02215, USA}
\address[addr2]{Condensed Matter Theory Group,
                Paul Scherrer Institut, CH-5232 Villigen PSI, Switzerland}
\address[addr3]{Laboratoire de Physique, {\'E}cole Normale Sup{\'e}rieure, 
                46 All{\'e}e d'Italie, 69364 Lyon Cedex 07, France}

\thanks[CC and CC]{
This work is supported in part by the NSF Grants DMR-0305482 and 
DMR-0403997.
       }

\begin{abstract}
We construct a local interacting quantum dimer model on the 
square lattice, whose 
zero-temperature phase diagram is characterized by a line of critical 
points separating two ordered phases of the valence bond crystal type. 
On one side, the line of critical points terminates in a 
quantum transition inherited from a Kosterlitz-Thouless transition 
in an associated classical model. 
We also discuss the effect of a longer-range dimer
interaction that can be used to suppress the line of critical points 
by gradually shrinking it to a single point. 
Finally, we propose a way to generalize the quantum Hamiltonian to a dilute 
dimer model in presence of monomers
and we qualitatively discuss the phase diagram. 
\end{abstract}


\end{frontmatter}
%
%

\section{\label{sec: intro}
Introduction
        }
Dimer models are of interest to a variety of scientific disciplines
{}from chemistry to mathematics and physics. 
In chemistry, dimers are used, for example, to model molecules deposited 
on crystalline surfaces and to study their thermodynamic 
properties~\cite{Fowler1937}.  
In mathematics, dimers are often used to construct combinatorial and 
folding problems such as the domino tiling of a two-dimensional ($2D$) 
plane~\cite{Thurston1990}. 
In physics, dimer models have been elevated from problems in
classical statistical physics%
~\cite{%
Kasteleyn1961,%
Fisher1961,%
Kondev1995,%
Kondev1996,%
Fendley2002,%
Krauth2003,%
Huse2003,%
Alet2005%
       },
to problems in quantum statistical physics%
~\cite{%
Kivelson1987,%
Rokhsar1988,%
Levitov1990,%
Fradkin1991,%
Leung1996,%
Henley1997,%
Moessner2001,%
Chandra2001,%
Moessner2002,%
Ioselevich2002,%
Misguich2002,%
Sondhi2003,%
Moessner2003,%
Freedman2003,%
Hermele2004,%
Fradkin2004,%
Misguich2004,%
Henley2004,%
Ardonne2004,%
Ivanov2004,%
Misguich2005,%
Ralko2005,%
Castelnovo2005,%
Raman2005,%
Freedman2005a,%
Syljuasen2005,%
Sandvik2005,%
Moessner2005,%
Bergman2005%
       },
with the advent of high-$T^{\ }_{\mathrm{c}}$ superconductivity. 
In particular, quantum dimer models can provide examples 
of strongly correlated quantum systems for which the zero temperature
phase diagram is characterized by exotic quantum phase transitions
that fall out of the classification of phase transitions
proposed by Landau~\cite{Senthil2004,Balents2005}. 
Furthermore, the finite-temperature phase diagram of quantum dimer models 
might give some insight into the phenomenological observation that scaling 
laws extend to surprisingly high temperatures in some strongly correlated 
systems~\cite{Chakravarty02,Castelnovo2006}.

In this paper, we show how dimer models can be used as a
laboratory to construct quantum Hamiltonians displaying
phase transitions that cannot be understood in terms of a local order
parameter, i.e., phase transitions that cannot be encoded by an
effective theory of the Landau-Ginzburg type, a topic of renewed interest in 
condensed matter physics~\cite{Senthil2004,Balents2005}.
Perhaps the most famous counter example
to a phase transition described with
a Landau-Ginzburg action for a local order parameter is
the Kosterlitz-Thouless (KT) transition. 
The KT transition is a weak essential singularity of the free energy
for a phase-like order parameter with support in $2D$
Euclidean space. It is interpreted as
the unbinding of topological defects (vortices)
in the order parameter. The main result of this paper is the construction of 
a quantum dimer model with {\it local interactions} on the square 
lattice~[Eqs.~(\ref{eq: def H0},\ref{eq: quantum Alet Hamiltonian})] 
that undergoes a quantum phase transition
of the KT type when measured by the
spatial decay of equal-time correlation functions.

It is well known that quantum phase transitions can be of the KT type
in $1D$ systems with dynamical exponent $z=1$ relating
the scaling in space to the scaling in time. 
For example, a $1D$ Luttinger liquid can be unstable to a 
charge-ordered density wave through a KT transition. 
This is so because the quantum field theory describing
the quantum phase transition for interacting fermions is related 
through bosonization to a scalar field theory, the Sine-Gordon model. 
Analytical continuation of time to imaginary time can be used to turn
(Minkowski) space-time into $2D$ Euclidean space 
while the Poincar\'e symmetry group becomes symmetry under translations
and rotations. Evidently, if a quantum phase transition
can be described by a local quantum field theory in $D+1$ space-time
that turns into a local classical action undergoing 
a classical phase transition in $D+1$ Euclidean space
upon analytical continuation of time to imaginary 
time~\cite{Hertz1976},
then this quantum phase transition cannot be associated to 
a KT transition when $z=1$ and $D\geq2$. Viewed against this no-go theorem, 
it is remarkable that some equal-time correlation functions of
a $2D$ quantum dimer model with local interactions and defined on the square 
lattice~[Eqs.~(\ref{eq: def H0},\ref{eq: quantum Alet Hamiltonian})] 
share the hallmarks of a KT transition.

Since the quantum Hamiltonian of our $2D$ lattice model, defined in 
Eqs.~(\ref{eq: def H0}) and~(\ref{eq: quantum Alet Hamiltonian}),
has only local interactions, it is safe to argue that unequal-time
correlations should show algebraic behavior if the equal-time
correlations do so. The reason is that 
a local Hamiltonian with algebraic spatial correlations 
should be gapless. A rigorous proof of this statement 
was given by Hastings in Ref.~\onlinecite{Hastings2004}. 
The converse statement is known not to be true, 
as shown in Ref.~\cite{Freedman2005b}. 
A local quantum Hamiltonian may be gapless but have only
short-ranged spatial correlations between local operators. Therefore,
while in this paper we concentrate solely on equal-time correlation
functions, the KT-like transition identified through the algebraic
spatial correlations should be manifest in unequal-time correlation
functions related to the spatial ones through a dynamical exponent
$z$. Moreover, a dynamical exponent $z=2$ is suggested by our mapping of 
the quantum system onto a classical model with local stochastic 
dynamics~\cite{Henley1997,Henley2004,Castelnovo2005}.

What is the continuum imaginary-time field theory that captures
the low-energy physics of the local quantum lattice Hamiltonian,
Eqs.~(\ref{eq: def H0},\ref{eq: quantum Alet Hamiltonian}),
when fine-tuned to its line of critical points?
We expect it to be that of a classical Lifshitz point problem 
in a uniform magnetic field introduced by Grinstein in 
Ref.~\onlinecite{Grinstein}. 
Here, a short-ranged anisotropic coupling between $2D$ layers, 
each of which are described by a local classical Lagrangian,
is interpreted as the coupling between imaginary-time
slices of the quantum problem~\cite{Ardonne2004,Ghaemi2005}. 
Note, however, that it is not always the case that a \emph{local} 
classical Lagrangian in $D+1$ Euclidean space corresponds to 
a \emph{local} quantum Hamiltonian in $D$ spatial dimensions; 
a counter-example was shown in Ref.~\onlinecite{Shtengel2005}. Our $2D$
lattice realization of the quantum KT-transition starts directly from
the local quantum Hamiltonian, and avoids any discussion of the
corresponding classical imaginary-time Lagrangian.

The quantum dimer model on the square lattice discussed in the present paper, 
Eqs.~(\ref{eq: def H0},\ref{eq: quantum Alet Hamiltonian}), 
is represented by a symmetric and positive matrix 
that obeys the so-called Stochastic Matrix Form (SMF) decomposition%
~\onlinecite{%
Henley2004,
Ardonne2004,%
Castelnovo2005%
             }.
The advantages of an SMF decomposition of a quantum Hamiltonian 
are three-fold.
First, at least one ground state (GS) can be obtained exactly 
in terms of the parameters entering the SMF quantum Hamiltonian,
see Eq.~(\ref{eq: quantum Alet GS})%
~\cite{%
Rokhsar1988,%
Henley1997,%
Henley2004,%
Ardonne2004,%
Castelnovo2005%
       }.
Second, one can construct a classical configuration space that is in 
one-to-one correspondence with the orthonormal basis in which the SMF 
quantum Hamiltonian is represented. 
On this configuration space, a classical partition function
can be uniquely defined from the GS wavefunction, 
see Eq.~(\ref{eq: Alet partition function})%
~\cite{%
Rokhsar1988,%
Henley1997,%
Henley2004,%
Ardonne2004,%
Castelnovo2005%
      }, 
such that zero-temperature and equal-time correlation functions of quantum 
operators diagonal in the SMF basis are equivalent to equilibrium thermal 
averages of corresponding quantities in the classical system. 
Note that the possibility to use classical numerical techniques, such as 
Monte Carlo simulations and transfer matrix calculations, in the 
\emph{same number of dimensions} gives access to much larger system 
sizes than quantum techniques, such as quantum Monte Carlo or 
exact diagonalization routines, do. 
Third, the parameters entering the SMF quantum Hamiltonian
allow us to define in a unique way the approach to equilibrium of the
associated classical system,
see Refs.~\onlinecite{Henley1997,Henley2004,Castelnovo2005}, i.e., 
in a way that zero-temperature, imaginary-time correlation functions of 
operators diagonal in the SMF basis can be obtained from real-time 
correlation functions in the stochastic classical system. 
In particular, if the partition function of the associated 
classical system undergoes a KT transition upon
varying the quantum parameters entering the SMF quantum Hamiltonian,
so does the equal-time GS expectation value of operators diagonal
in the SMF basis. We can now understand how it is
possible to circumvent the no-go theorem. The no-go theorem assumes
that a classical phase transition faithfully represents
\textit{all} correlation functions in a quantum phase transition
with $z=1$. 
In this paper, only equal-time GS expectation values of operators
diagonal in the SMF basis are faithfully represented by correlation
functions at the KT critical point since the value of $z\neq1$ 
is not known rigorously. Similar results have been announced by 
Papanikolaou~\textit{et al.} in Ref~\onlinecite{Papanikolaou2006}.

The paper is organized as follows.
%
%
We will show in Sec.~\ref{sec: quantum model}
that the GS~(\ref{eq: quantum Alet GS}) 
\begin{equation}
\vert\Psi^{\ }_{0}\rangle 
= 
\sum_{\mathcal{C}\in\mathcal{S}^{\ }_{0}} 
  e^{\frac{u}{2T} N^{(f)}_{\mathcal{C}}} 
    \vert\mathcal{C}\rangle 
\nonumber
\end{equation}
of quantum Hamiltonian~(\ref{eq: def H0})
defines the classical partition function~(\ref{eq: Alet partition function}) 
for interacting dimers on the square lattice 
\begin{equation}
Z(T/u) 
:= 
\sum_{\mathcal{C}\in\mathcal{S}^{\ }_{0}} 
e^{-E^{(u)}_{\mathcal{C}}/T},
\qquad
E^{(u)}_{\mathcal{C}}:=-u N^{(f)}_{\mathcal{C}}. 
\nonumber
\end{equation} 
Here $\mathcal{S}^{\ }_{0}$ is the set of all possible classical dimer 
configurations on the square lattice, $u$ and $T$ are two real parameters, 
and $N^{(f)}_{\mathcal{C}}$ is the number of plaquettes having two parallel 
dimers in configuration $\mathcal{C}$. 
The classical partition function~(\ref{eq: Alet partition function}) 
was studied numerically by Alet~\textit{et al.} in Ref.~\onlinecite{Alet2005}
for one sign of the interaction between the dimers. 
In Sec.~\ref{sec: classical model} we extend the numerical study by 
Alet~\textit{et al.} to the other sign of the interaction between the dimers. 
The temperature of the classical partition 
function~(\ref{eq: Alet partition function}) plays the role of 
a quantum coupling in the quantum Hamiltonian~(\ref{eq: def H0}).
Ground-state equal-time expectation values of operators diagonal in
the dimer basis are thus inherited 
from the thermodynamics of the classical partition function. 
In the high-temperature regime, the associated classical system
exhibits a line of critical points.
As the temperature is lowered, the classical system undergoes 
either a first-order or KT transition
depending on the sign of the interactions between the dimers. 
%
%
To study the robustness of the line of critical points in the 
zero-temperature phase diagram of quantum Hamiltonian~(\ref{eq: def H0}),
we extend the range of the dimer interactions in 
Sec.~\ref{sec: long range interactions}.
We show that, for one sign of the longer-range dimer interaction,
the line of critical points shrinks continuously upon increasing the
strength of this longer-range interaction.
%
%
Section~\ref{sec: monomers}
is devoted to another kind of perturbation to our 
quantum Hamiltonian~(\ref{eq: def H0}): the presence of defects represented 
by sites not occupied by a dimer (monomers). 
We define the more general SMF quantum 
Hamiltonian~(\ref{eq: quantum Alet + monomer Hamiltonian}) 
that accounts for the existence of monomers 
in a dilute dimer model. 
Once again the GS~(\ref{eq: quantum Alet + monomer GS}) can be computed 
exactly 
\begin{equation}
\vert\Psi^{\ }_{\textrm{\small tot}}\rangle 
= 
\sum_{
\mathcal{C}\in
\mathcal{S}^{\ }_{} 
     } 
  e^{(u N^{(f)}_{\mathcal{C}} + \mu M^{\ }_{\mathcal{C}})/2T} 
    \vert\mathcal{C}\rangle, 
\nonumber
\end{equation}
where, in addition to the quantities defined above, 
$\mathcal{S}$ is the set of all classical dilute dimer 
configurations on the square lattice (i.e., where each site belongs to 
\emph{at most} one dimer), $\mu$ is the chemical potential for monomers, 
and $M^{\ }_{\mathcal{C}}$ is the total number of monomers in configuration 
$\mathcal{C}$. 
We can then establish a correspondence between our quantum SMF 
Hamiltonian and a classical dilute dimer model described by the partition 
function~(\ref{eq: Alet + monomer partition function})
\begin{equation}
Z(T/u,\mu/T) 
:= 
\sum_{
\mathcal{C}\in
\mathcal{S}^{\ }_{}
     }
  \exp \left(-\frac{E^{(u,\mu)}_{\mathcal{C}}}{T} \right) 
= 
\sum_{
\mathcal{C}\in
\mathcal{S}^{\ }_{}
     }
  \exp 
    \left(
      \frac{u N^{(f)}_{\mathcal{C}} + \mu M^{\ }_{\mathcal{C}}}{T} 
    \right). 
\nonumber
\end{equation}
This is a promising result as equal-time correlation functions for monomers 
can thereby be studied, for example, using classical Monte Carlo 
algorithms~\cite{Alet2005}. 
Of course, the computation of unequal-time correlation functions
for monomers still requires the use of quantum Monte Carlo simulations
or exact diagonalization techniques and is therefore limited to smaller system 
sizes. 
%
%

\section{\label{sec: quantum model}
A square lattice interacting quantum dimer model with solvable ground state
        }

We begin our construction of a quantum square lattice dimer model
(SLDM) whose ground state (GS) is exactly solvable by revisiting the
quantum dimer model that was introduced by Rokhsar and Kivelson (RK) in
Ref.~\onlinecite{Rokhsar1988}. The Hilbert space is given by the span
of the orthonormal basis states $\vert\mathcal{C}\rangle$ labeled by
all the classical dimer configurations $\mathcal{C}$ of the square
lattice. A classical dimer configuration on the square lattice is
obtained by covering all the bonds connecting nearest-neighbor sites
with dimers in such a way that each site is the end point of one and
only one dimer. The set off all allowed dimer coverings will be
denoted by $\mathcal{S}^{\ }_{0}$. 

The RK Hamiltonian acting on this Hilbert space is then commonly written 
as 
%
%
\begin{equation}
\widehat{H}^{\ }_{RK} 
= 
\sum^{\ }_{p} 
  \left[ \vphantom{\Big[} 
  v \left( \vphantom{\Big[} 
      \vert
\setlength{\unitlength}{0.06mm}
\begin{picture}(90,55)(0,8) 
  \linethickness{0.04mm}
  \put(18,1){\line(1,0){54}} 
  \put(18,1){\circle*{20}}
  \linethickness{0.5mm}
  \put(72,1){\line(0,1){54}}
  \put(72,1){\circle*{20}}
  \linethickness{0.04mm}
  \put(72,55){\line(-1,0){54}}
  \put(72,55){\circle*{20}}
  \linethickness{0.5mm}
  \put(18,55){\line(0,-1){54}}
  \put(18,55){\circle*{20}}
\end{picture}
\rangle \langle
\setlength{\unitlength}{0.06mm}
\begin{picture}(90,55)(0,8) 
  \linethickness{0.04mm}
  \put(18,1){\line(1,0){54}} 
  \put(18,1){\circle*{20}}
  \linethickness{0.5mm}
  \put(72,1){\line(0,1){54}}
  \put(72,1){\circle*{20}}
  \linethickness{0.04mm}
  \put(72,55){\line(-1,0){54}}
  \put(72,55){\circle*{20}}
  \linethickness{0.5mm}
  \put(18,55){\line(0,-1){54}}
  \put(18,55){\circle*{20}}
\end{picture}
\vert_p
  + 
      \vert
\setlength{\unitlength}{0.06mm}
\begin{picture}(90,55)(0,8) 
  \linethickness{0.5mm}
  \put(18,1){\line(1,0){54}} 
  \put(18,1){\circle*{20}}
  \linethickness{0.04mm}
  \put(72,1){\line(0,1){54}}
  \put(72,1){\circle*{20}}
  \linethickness{0.5mm}
  \put(72,55){\line(-1,0){54}}
  \put(72,55){\circle*{20}}
  \linethickness{0.04mm}
  \put(18,55){\line(0,-1){54}}
  \put(18,55){\circle*{20}}
\end{picture}
\rangle \langle
\setlength{\unitlength}{0.06mm}
\begin{picture}(90,55)(0,8) 
  \linethickness{0.5mm}
  \put(18,1){\line(1,0){54}} 
  \put(18,1){\circle*{20}}
  \linethickness{0.04mm}
  \put(72,1){\line(0,1){54}}
  \put(72,1){\circle*{20}}
  \linethickness{0.5mm}
  \put(72,55){\line(-1,0){54}}
  \put(72,55){\circle*{20}}
  \linethickness{0.04mm}
  \put(18,55){\line(0,-1){54}}
  \put(18,55){\circle*{20}}
\end{picture}
\vert_p
    \right) 
- t \left( \vphantom{\Big[} 
      \vert
\setlength{\unitlength}{0.06mm}
\begin{picture}(90,55)(0,8) 
  \linethickness{0.04mm}
  \put(18,1){\line(1,0){54}} 
  \put(18,1){\circle*{20}}
  \linethickness{0.5mm}
  \put(72,1){\line(0,1){54}}
  \put(72,1){\circle*{20}}
  \linethickness{0.04mm}
  \put(72,55){\line(-1,0){54}}
  \put(72,55){\circle*{20}}
  \linethickness{0.5mm}
  \put(18,55){\line(0,-1){54}}
  \put(18,55){\circle*{20}}
\end{picture}
\rangle \langle
\setlength{\unitlength}{0.06mm}
\begin{picture}(90,55)(0,8) 
  \linethickness{0.5mm}
  \put(18,1){\line(1,0){54}} 
  \put(18,1){\circle*{20}}
  \linethickness{0.04mm}
  \put(72,1){\line(0,1){54}}
  \put(72,1){\circle*{20}}
  \linethickness{0.5mm}
  \put(72,55){\line(-1,0){54}}
  \put(72,55){\circle*{20}}
  \linethickness{0.04mm}
  \put(18,55){\line(0,-1){54}}
  \put(18,55){\circle*{20}}
\end{picture}
\vert_p
  + 
      \vert
\setlength{\unitlength}{0.06mm}
\begin{picture}(90,55)(0,8) 
  \linethickness{0.5mm}
  \put(18,1){\line(1,0){54}} 
  \put(18,1){\circle*{20}}
  \linethickness{0.04mm}
  \put(72,1){\line(0,1){54}}
  \put(72,1){\circle*{20}}
  \linethickness{0.5mm}
  \put(72,55){\line(-1,0){54}}
  \put(72,55){\circle*{20}}
  \linethickness{0.04mm}
  \put(18,55){\line(0,-1){54}}
  \put(18,55){\circle*{20}}
\end{picture}
\rangle \langle
\setlength{\unitlength}{0.06mm}
\begin{picture}(90,55)(0,8) 
  \linethickness{0.04mm}
  \put(18,1){\line(1,0){54}} 
  \put(18,1){\circle*{20}}
  \linethickness{0.5mm}
  \put(72,1){\line(0,1){54}}
  \put(72,1){\circle*{20}}
  \linethickness{0.04mm}
  \put(72,55){\line(-1,0){54}}
  \put(72,55){\circle*{20}}
  \linethickness{0.5mm}
  \put(18,55){\line(0,-1){54}}
  \put(18,55){\circle*{20}}
\end{picture}
\vert_p
    \right) 
  \right] 
\label{eq: RK Hamiltonian}
\end{equation}
where the summation is over all plaquettes $p$ of the square lattice.
The operators 
$\vert
\setlength{\unitlength}{0.06mm}
\begin{picture}(90,55)(0,8) 
  \linethickness{0.04mm}
  \put(18,1){\line(1,0){54}} 
  \put(18,1){\circle*{20}}
  \linethickness{0.5mm}
  \put(72,1){\line(0,1){54}}
  \put(72,1){\circle*{20}}
  \linethickness{0.04mm}
  \put(72,55){\line(-1,0){54}}
  \put(72,55){\circle*{20}}
  \linethickness{0.5mm}
  \put(18,55){\line(0,-1){54}}
  \put(18,55){\circle*{20}}
\end{picture}
\rangle \langle
\setlength{\unitlength}{0.06mm}
\begin{picture}(90,55)(0,8) 
  \linethickness{0.04mm}
  \put(18,1){\line(1,0){54}} 
  \put(18,1){\circle*{20}}
  \linethickness{0.5mm}
  \put(72,1){\line(0,1){54}}
  \put(72,1){\circle*{20}}
  \linethickness{0.04mm}
  \put(72,55){\line(-1,0){54}}
  \put(72,55){\circle*{20}}
  \linethickness{0.5mm}
  \put(18,55){\line(0,-1){54}}
  \put(18,55){\circle*{20}}
\end{picture}
\vert_p$ 
and 
$\vert
\setlength{\unitlength}{0.06mm}
\begin{picture}(90,55)(0,8) 
  \linethickness{0.5mm}
  \put(18,1){\line(1,0){54}} 
  \put(18,1){\circle*{20}}
  \linethickness{0.04mm}
  \put(72,1){\line(0,1){54}}
  \put(72,1){\circle*{20}}
  \linethickness{0.5mm}
  \put(72,55){\line(-1,0){54}}
  \put(72,55){\circle*{20}}
  \linethickness{0.04mm}
  \put(18,55){\line(0,-1){54}}
  \put(18,55){\circle*{20}}
\end{picture}
\rangle \langle
\setlength{\unitlength}{0.06mm}
\begin{picture}(90,55)(0,8) 
  \linethickness{0.5mm}
  \put(18,1){\line(1,0){54}} 
  \put(18,1){\circle*{20}}
  \linethickness{0.04mm}
  \put(72,1){\line(0,1){54}}
  \put(72,1){\circle*{20}}
  \linethickness{0.5mm}
  \put(72,55){\line(-1,0){54}}
  \put(72,55){\circle*{20}}
  \linethickness{0.04mm}
  \put(18,55){\line(0,-1){54}}
  \put(18,55){\circle*{20}}
\end{picture}
\vert_p$
denote projection operators onto the subspace of states associated
with configurations $\mathcal{C}$ that contain two parallel dimers
(vertical or horizontal, respectively) on plaquette $p$.
The operators $\vert
\setlength{\unitlength}{0.06mm}
\begin{picture}(90,55)(0,8) 
  \linethickness{0.04mm}
  \put(18,1){\line(1,0){54}} 
  \put(18,1){\circle*{20}}
  \linethickness{0.5mm}
  \put(72,1){\line(0,1){54}}
  \put(72,1){\circle*{20}}
  \linethickness{0.04mm}
  \put(72,55){\line(-1,0){54}}
  \put(72,55){\circle*{20}}
  \linethickness{0.5mm}
  \put(18,55){\line(0,-1){54}}
  \put(18,55){\circle*{20}}
\end{picture}
\rangle
\langle
\setlength{\unitlength}{0.06mm}
\begin{picture}(90,55)(0,8) 
  \linethickness{0.5mm}
  \put(18,1){\line(1,0){54}} 
  \put(18,1){\circle*{20}}
  \linethickness{0.04mm}
  \put(72,1){\line(0,1){54}}
  \put(72,1){\circle*{20}}
  \linethickness{0.5mm}
  \put(72,55){\line(-1,0){54}}
  \put(72,55){\circle*{20}}
  \linethickness{0.04mm}
  \put(18,55){\line(0,-1){54}}
  \put(18,55){\circle*{20}}
\end{picture}
\vert_p$ and
$\vert
\setlength{\unitlength}{0.06mm}
\begin{picture}(90,55)(0,8) 
  \linethickness{0.5mm}
  \put(18,1){\line(1,0){54}} 
  \put(18,1){\circle*{20}}
  \linethickness{0.04mm}
  \put(72,1){\line(0,1){54}}
  \put(72,1){\circle*{20}}
  \linethickness{0.5mm}
  \put(72,55){\line(-1,0){54}}
  \put(72,55){\circle*{20}}
  \linethickness{0.04mm}
  \put(18,55){\line(0,-1){54}}
  \put(18,55){\circle*{20}}
\end{picture}
\rangle
\langle
\setlength{\unitlength}{0.06mm}
\begin{picture}(90,55)(0,8) 
  \linethickness{0.04mm}
  \put(18,1){\line(1,0){54}} 
  \put(18,1){\circle*{20}}
  \linethickness{0.5mm}
  \put(72,1){\line(0,1){54}}
  \put(72,1){\circle*{20}}
  \linethickness{0.04mm}
  \put(72,55){\line(-1,0){54}}
  \put(72,55){\circle*{20}}
  \linethickness{0.5mm}
  \put(18,55){\line(0,-1){54}}
  \put(18,55){\circle*{20}}
\end{picture}
\vert_p$ are the plaquette $p$ flipping
operators that maps any state $\vert\mathcal{C}\rangle$ with two
horizontal or vertical dimers at plaquette $p$ onto the state
$\vert\overline{\mathcal{C}}\rangle$ obtained by rotating the two
parallel dimers at plaquette $p$ by $90$ degrees, while it annihilates
state $\vert\mathcal{C}\rangle$ otherwise (i.e., if the
configuration $\mathcal{C}$ does not contain two horizontal or vertical 
parallel dimers at plaquette $p$). 
Given any classical dimer configuration $\mathcal{C}$,
we shall call any plaquette occupied by two parallel dimers a
flippable plaquette.

The representation~(\ref{eq: RK Hamiltonian}) of the RK
Hamiltonian is explicitly local as no reference is made to the dimer
covering away from plaquette $p$ in the definitions of operators
$\vert
\setlength{\unitlength}{0.06mm}
\begin{picture}(90,55)(0,8) 
  \linethickness{0.04mm}
  \put(18,1){\line(1,0){54}} 
  \put(18,1){\circle*{20}}
  \linethickness{0.5mm}
  \put(72,1){\line(0,1){54}}
  \put(72,1){\circle*{20}}
  \linethickness{0.04mm}
  \put(72,55){\line(-1,0){54}}
  \put(72,55){\circle*{20}}
  \linethickness{0.5mm}
  \put(18,55){\line(0,-1){54}}
  \put(18,55){\circle*{20}}
\end{picture}
\rangle
\langle
\setlength{\unitlength}{0.06mm}
\begin{picture}(90,55)(0,8) 
  \linethickness{0.04mm}
  \put(18,1){\line(1,0){54}} 
  \put(18,1){\circle*{20}}
  \linethickness{0.5mm}
  \put(72,1){\line(0,1){54}}
  \put(72,1){\circle*{20}}
  \linethickness{0.04mm}
  \put(72,55){\line(-1,0){54}}
  \put(72,55){\circle*{20}}
  \linethickness{0.5mm}
  \put(18,55){\line(0,-1){54}}
  \put(18,55){\circle*{20}}
\end{picture}
\vert_p$,
$\vert
\setlength{\unitlength}{0.06mm}
\begin{picture}(90,55)(0,8) 
  \linethickness{0.5mm}
  \put(18,1){\line(1,0){54}} 
  \put(18,1){\circle*{20}}
  \linethickness{0.04mm}
  \put(72,1){\line(0,1){54}}
  \put(72,1){\circle*{20}}
  \linethickness{0.5mm}
  \put(72,55){\line(-1,0){54}}
  \put(72,55){\circle*{20}}
  \linethickness{0.04mm}
  \put(18,55){\line(0,-1){54}}
  \put(18,55){\circle*{20}}
\end{picture}
\rangle \langle
\setlength{\unitlength}{0.06mm}
\begin{picture}(90,55)(0,8) 
  \linethickness{0.5mm}
  \put(18,1){\line(1,0){54}} 
  \put(18,1){\circle*{20}}
  \linethickness{0.04mm}
  \put(72,1){\line(0,1){54}}
  \put(72,1){\circle*{20}}
  \linethickness{0.5mm}
  \put(72,55){\line(-1,0){54}}
  \put(72,55){\circle*{20}}
  \linethickness{0.04mm}
  \put(18,55){\line(0,-1){54}}
  \put(18,55){\circle*{20}}
\end{picture}
\vert_p$ and
$\vert
\setlength{\unitlength}{0.06mm}
\begin{picture}(90,55)(0,8) 
  \linethickness{0.04mm}
  \put(18,1){\line(1,0){54}} 
  \put(18,1){\circle*{20}}
  \linethickness{0.5mm}
  \put(72,1){\line(0,1){54}}
  \put(72,1){\circle*{20}}
  \linethickness{0.04mm}
  \put(72,55){\line(-1,0){54}}
  \put(72,55){\circle*{20}}
  \linethickness{0.5mm}
  \put(18,55){\line(0,-1){54}}
  \put(18,55){\circle*{20}}
\end{picture}
\rangle
\langle
\setlength{\unitlength}{0.06mm}
\begin{picture}(90,55)(0,8) 
  \linethickness{0.5mm}
  \put(18,1){\line(1,0){54}} 
  \put(18,1){\circle*{20}}
  \linethickness{0.04mm}
  \put(72,1){\line(0,1){54}}
  \put(72,1){\circle*{20}}
  \linethickness{0.5mm}
  \put(72,55){\line(-1,0){54}}
  \put(72,55){\circle*{20}}
  \linethickness{0.04mm}
  \put(18,55){\line(0,-1){54}}
  \put(18,55){\circle*{20}}
\end{picture}
\vert_p$,
$\vert
\setlength{\unitlength}{0.06mm}
\begin{picture}(90,55)(0,8) 
  \linethickness{0.5mm}
  \put(18,1){\line(1,0){54}} 
  \put(18,1){\circle*{20}}
  \linethickness{0.04mm}
  \put(72,1){\line(0,1){54}}
  \put(72,1){\circle*{20}}
  \linethickness{0.5mm}
  \put(72,55){\line(-1,0){54}}
  \put(72,55){\circle*{20}}
  \linethickness{0.04mm}
  \put(18,55){\line(0,-1){54}}
  \put(18,55){\circle*{20}}
\end{picture}
\rangle \langle
\setlength{\unitlength}{0.06mm}
\begin{picture}(90,55)(0,8) 
  \linethickness{0.04mm}
  \put(18,1){\line(1,0){54}} 
  \put(18,1){\circle*{20}}
  \linethickness{0.5mm}
  \put(72,1){\line(0,1){54}}
  \put(72,1){\circle*{20}}
  \linethickness{0.04mm}
  \put(72,55){\line(-1,0){54}}
  \put(72,55){\circle*{20}}
  \linethickness{0.5mm}
  \put(18,55){\line(0,-1){54}}
  \put(18,55){\circle*{20}}
\end{picture}
\vert_p$.
The nature of the GS of the RK Hamiltonian depends on the
dimensionless ratio between the characteristic energies
$v\in\mathbb{R}$ and $t\geq0$.  When $v/t\ll-1$, the GS is expected to
display columnar ordering of the dimers, i.e., each dimer has
precisely two parallel neighboring dimers.  When $v/t\gg+1$, the GS is
expected to display a staggered ordering of the dimers, i.e., no two
parallel neighboring dimers are present in the system.  The so-called
RK (critical) point
\begin{equation}
v=t
\label{eq: def RK point}
\end{equation}
in parameter space is special in that the GS is known exactly%
~\cite{Rokhsar1988}
\begin{equation}
\vert\Psi\rangle:=
\sum_{\mathcal{C}\in\mathcal{S}^{\ }_{0}}
\vert\mathcal{C}\rangle
\label{eq: RK GS}
\end{equation} 
and it is non-degenerate within each irreducible sector of the Hilbert space 
under the action of the RK Hamiltonian. It separates two ordered phases%
~\cite{Syljuasen2005}.

The notation used so far is not suitable to construct generalizations
of the RK Hamiltonian.  What we need instead is a representation that
will allow us to ``decorate'' the plaquettes so as to encode the
effects of interactions between dimers beyond just plaquettes with
parallel or non-parallel dimers. To this end, let us begin by
rewriting the RK Hamiltonian in terms of local objects
$\ell^{(p)}_{0}$ that denote one of the two possible flipppable dimer
configurations on plaquette $p$, i.e., either
$\ell^{(p)}_{0}=
\setlength{\unitlength}{0.06mm}
\begin{picture}(90,55)(0,8) 
  \linethickness{0.04mm}
  \put(18,1){\line(1,0){54}} 
  \put(18,1){\circle*{20}}
  \linethickness{0.5mm}
  \put(72,1){\line(0,1){54}}
  \put(72,1){\circle*{20}}
  \linethickness{0.04mm}
  \put(72,55){\line(-1,0){54}}
  \put(72,55){\circle*{20}}
  \linethickness{0.5mm}
  \put(18,55){\line(0,-1){54}}
  \put(18,55){\circle*{20}}
\end{picture}
_p$ or
$\ell^{(p)}_{0}=
\setlength{\unitlength}{0.06mm}
\begin{picture}(90,55)(0,8) 
  \linethickness{0.5mm}
  \put(18,1){\line(1,0){54}} 
  \put(18,1){\circle*{20}}
  \linethickness{0.04mm}
  \put(72,1){\line(0,1){54}}
  \put(72,1){\circle*{20}}
  \linethickness{0.5mm}
  \put(72,55){\line(-1,0){54}}
  \put(72,55){\circle*{20}}
  \linethickness{0.04mm}
  \put(18,55){\line(0,-1){54}}
  \put(18,55){\circle*{20}}
\end{picture}
_p$ (more formally,
$\ell^{(p)}_{0}\in\{
\setlength{\unitlength}{0.06mm}
\begin{picture}(90,55)(0,8) 
  \linethickness{0.04mm}
  \put(18,1){\line(1,0){54}} 
  \put(18,1){\circle*{20}}
  \linethickness{0.5mm}
  \put(72,1){\line(0,1){54}}
  \put(72,1){\circle*{20}}
  \linethickness{0.04mm}
  \put(72,55){\line(-1,0){54}}
  \put(72,55){\circle*{20}}
  \linethickness{0.5mm}
  \put(18,55){\line(0,-1){54}}
  \put(18,55){\circle*{20}}
\end{picture}
_p,
\setlength{\unitlength}{0.06mm}
\begin{picture}(90,55)(0,8) 
  \linethickness{0.5mm}
  \put(18,1){\line(1,0){54}} 
  \put(18,1){\circle*{20}}
  \linethickness{0.04mm}
  \put(72,1){\line(0,1){54}}
  \put(72,1){\circle*{20}}
  \linethickness{0.5mm}
  \put(72,55){\line(-1,0){54}}
  \put(72,55){\circle*{20}}
  \linethickness{0.04mm}
  \put(18,55){\line(0,-1){54}}
  \put(18,55){\circle*{20}}
\end{picture}
_p\}$). Given
the flippable plaquette $\ell^{(p)}_{0}$ we denote by
$\overline{\ell^{(p)}_{0}}$ the flippable plaquette $p$ obtained from a
$90^{\textrm{\scriptsize o}}$ rotation of the dimer covering in
$\ell^{(p)}_{0}$ (i.e., if
$\ell^{(p)}_{0}=
\setlength{\unitlength}{0.06mm}
\begin{picture}(90,55)(0,8) 
  \linethickness{0.04mm}
  \put(18,1){\line(1,0){54}} 
  \put(18,1){\circle*{20}}
  \linethickness{0.5mm}
  \put(72,1){\line(0,1){54}}
  \put(72,1){\circle*{20}}
  \linethickness{0.04mm}
  \put(72,55){\line(-1,0){54}}
  \put(72,55){\circle*{20}}
  \linethickness{0.5mm}
  \put(18,55){\line(0,-1){54}}
  \put(18,55){\circle*{20}}
\end{picture}
_p$ then
$\overline{\ell^{(p)}_{0}}=
\setlength{\unitlength}{0.06mm}
\begin{picture}(90,55)(0,8) 
  \linethickness{0.5mm}
  \put(18,1){\line(1,0){54}} 
  \put(18,1){\circle*{20}}
  \linethickness{0.04mm}
  \put(72,1){\line(0,1){54}}
  \put(72,1){\circle*{20}}
  \linethickness{0.5mm}
  \put(72,55){\line(-1,0){54}}
  \put(72,55){\circle*{20}}
  \linethickness{0.04mm}
  \put(18,55){\line(0,-1){54}}
  \put(18,55){\circle*{20}}
\end{picture}
_p$, or if
$\ell^{(p)}_{0}=
\setlength{\unitlength}{0.06mm}
\begin{picture}(90,55)(0,8) 
  \linethickness{0.5mm}
  \put(18,1){\line(1,0){54}} 
  \put(18,1){\circle*{20}}
  \linethickness{0.04mm}
  \put(72,1){\line(0,1){54}}
  \put(72,1){\circle*{20}}
  \linethickness{0.5mm}
  \put(72,55){\line(-1,0){54}}
  \put(72,55){\circle*{20}}
  \linethickness{0.04mm}
  \put(18,55){\line(0,-1){54}}
  \put(18,55){\circle*{20}}
\end{picture}
_p$ then
$\overline{\ell^{(p)}_{0}}=
\setlength{\unitlength}{0.06mm}
\begin{picture}(90,55)(0,8) 
  \linethickness{0.04mm}
  \put(18,1){\line(1,0){54}} 
  \put(18,1){\circle*{20}}
  \linethickness{0.5mm}
  \put(72,1){\line(0,1){54}}
  \put(72,1){\circle*{20}}
  \linethickness{0.04mm}
  \put(72,55){\line(-1,0){54}}
  \put(72,55){\circle*{20}}
  \linethickness{0.5mm}
  \put(18,55){\line(0,-1){54}}
  \put(18,55){\circle*{20}}
\end{picture}
_p$).

The RK Hamiltonian then becomes
\begin{equation}
\widehat{H}^{\ }_{RK} 
= 
\frac{1}{2} \sum^{\ }_{\ell^{(p)}_{0}}
\widehat{Q}^{\ }_{\ell^{(p)}_{0}}
\label{eq: RK Hamiltonian a}
\end{equation}
where the sum over the local operators
\begin{equation}
\widehat{Q}^{\ }_{\ell^{(p)}_{0}}
:= 
  v \left( \vphantom{\Big[} 
      \vert\ell^{(p)}_{0}\rangle 
      \langle\ell^{(p)}_{0}\vert 
  + 
      \vert\overline{\ell^{(p)}_{0}}\rangle 
      \langle\overline{\ell^{(p)}_{0}}\vert
    \right) 
- t \left( \vphantom{\Big[} 
      \vert\ell^{(p)}_{0}\rangle 
      \langle\overline{\ell^{(p)}_{0}}\vert 
  + 
      \vert\overline{\ell^{(p)}_{0}}\rangle 
      \langle\ell^{(p)}_{0}\vert 
    \right)
\label{eq: RK Hamiltonian b gen}
\end{equation}
runs over all flippable plaquettes.
{}From now on we will drop the explicit dependence on the plaquette index
$p$ in the notation for a flippable plaquette: 
$\ell^{(p)}_{0}\equiv\ell^{\ }_{0}$.

With this notation in hand, one can extend the definition of a
flippable plaquette $\ell^{\ }_{0}$ to that of a \textit{decorated}
flippable plaquette $\ell^{*}_{0}$ in the following way.  Whereas
$\ell^{\ }_{0}$ specifies the position of the plaquette and the
orientation of the two parallel dimers covering it, $\ell^{*}_{0}$
contains additional information on the configuration of dimers at
neighboring plaquettes. This information can be encoded in a vector
$\mathbf{m}\in\{0,1\}^{n}$ that lists whether each of $n$ neighboring
bonds is occupied by a dimer or not. For example, $\mathbf{m} = (m^{\
}_{1},\,\ldots,\,m^{\ }_{4})$ when the additional information
corresponds to specifying whether the four bonds that face the four
edges of the plaquette $p$ are occupied by dimers ($m^{\ }_{i}=1$) or
not ($m^{\ }_{i}=0$), as is illustrated in Fig.~\ref{fig: decorated
plaquette}.
\begin{figure}[ht]
\centering
\includegraphics[width=0.8\columnwidth]{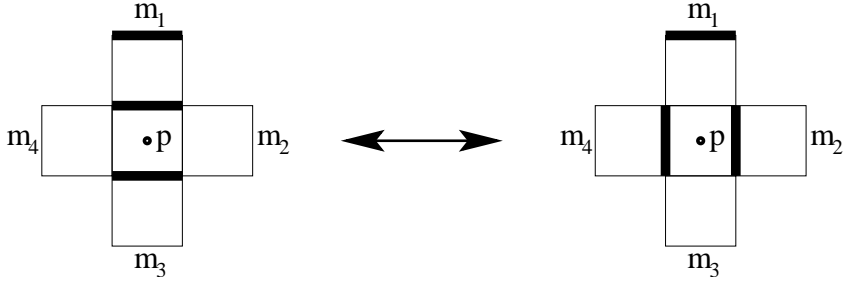}
\caption{
\label{fig: decorated plaquette}
Example of a local update 
$\ell^{*}_{0} \leftrightarrow \overline{\ell^{*}_{0}}$ 
between decorated flippable plaquettes. 
A decorated flippable plaquette is defined by the choice of the plaquette 
position $p$ on the square lattice, by 
the orientation of the two parallel dimers covering the plaquette $p$
(e.g., horizontal for $\ell^{*}_{0}$ and vertical for 
$\overline{\ell^{*}_{0}}$ in this example), and by the values $0$ or $1$ 
taken by the four parameters $m^{\ }_{i}$, $i=1,\,\ldots,\,4$ 
defined on the four bonds that face the four edges
of plaquette $p$.
The value $m^{\ }_{i}=1$ ($m^{\ }_{i}=0$) corresponds to bond $i$ 
being covered (not covered) by a dimer. 
In this example 
$\mathbf{m} = (1,0,0,0)$.
        }
\end{figure}
Given a flippable plaquette $\ell^{*}$,
the operation $\ell^{*}_{0}\to\overline{\ell^{*}_{0}}$
is defined by flipping the flippable plaquette $p$
(see Fig.~\ref{fig: decorated plaquette}).
As before, to any decorated flippable plaquette $\ell^{*}_{0}$
corresponds one and only one $\overline{\ell^{*}_{0}}$.
A straightforward generalization of the RK Hamiltonian
(\ref{eq: RK Hamiltonian a}) 
then follows by replacing the sum over flippable plaquettes with 
the sum over decorated flippable plaquettes,
\begin{equation}
\widehat{H}
= 
\frac{1}{2} \sum^{\ }_{\ell^{*}_{0}}
\widehat{Q}^{\ }_{\ell^{*}_{0}},
\label{eq: Hamiltonian a}
\end{equation}
where 
\begin{equation}
\widehat{Q}^{\ }_{\ell^{*}_{0}}
:= 
  v(\mathbf{m})\,
      \vert\ell^{*}_{0}\rangle 
      \langle\ell^{*}_{0}\vert 
  + 
  \overline{v(\mathbf{m})}\,
      \vert\overline{\ell^{*}_{0}}\rangle 
      \langle\overline{\ell^{*}_{0}}\vert
  - 
  t(\mathbf{m})  
    \left( \vphantom{\Big[} 
      \vert\ell^{*}_{0}\rangle 
      \langle\overline{\ell^{*}_{0}}\vert 
  + 
      \vert\overline{\ell^{*}_{0}}\rangle 
      \langle\ell^{*}_{0}\vert 
    \right).
\label{eq: Hamiltonian b gen}
\end{equation}
%
%
%
This Hamiltonian remains local although now the coupling constants 
$v(\mathbf{m}),\:\overline{v(\mathbf{m})}\in\mathbb{R}$ 
and
$t(\mathbf{m})\in\mathbb{R}$ can be used to 
encode interactions that extend beyond the two parallel dimers 
at a given flippable plaquette and involve, for example, the 
four dimers belonging to the four plaquettes
that share a bond with plaquette $p$ in the case 
of Fig.~\ref{fig: decorated plaquette}.
The GS for generic values
of $v(\mathbf{m}),\:\overline{v(\mathbf{m})}\in\mathbb{R}$
and
$t(\mathbf{m})\in\mathbb{R}$
is not known in closed form.
However, when
\begin{equation}
v(\mathbf{m}) \, \overline{v(\mathbf{m})} 
= 
t^2(\mathbf{m}),
\qquad
t(\mathbf{m})>0,
\end{equation}
the GS is given by
\begin{equation}
\vert\Psi\rangle:=
\sum_{\mathcal{C}\in\mathcal{S}^{\ }_{0}}
C(\mathcal{C})
\vert\mathcal{C}\rangle
\label{eq: GS general}
\end{equation}
within any irreducible sector of the Hilbert space under the action of 
the RK Hamiltonian provided the following integrability condition relating
the local data $v(\mathbf{m})$ and the global expansion coefficient 
$C(\mathcal{C})$ holds~\cite{Castelnovo2005}. 
For any dimer configuration $\mathcal{C}$ 
and for any decorated flippable plaquette $\ell^{*}_{0}$ 
present in $\mathcal{C}$ the dimer configuration, $\overline{\mathcal{C}}$ 
is uniquely defined by replacing $\ell^{*}_{0}$ with 
$\overline{\ell^{*}_{0}}$, 
while the integrability condition is satisfied whenever
\begin{equation}
C(\overline{\mathcal{C}})/C(\mathcal{C})=
\frac{1}{2}
\left(
v(\mathbf{m}) / t(\mathbf{m})
+
t(\mathbf{m}) / \overline{v(\mathbf{m})}
\right)
\label{eq: integrability condition}
\end{equation}
holds. 
The conditions above precisely define the Stochastic Matrix Form 
(SMF) decomposition of a quantum Hamiltonian discussed in 
Sec.~\ref{sec: intro}. 
For the remaining of this section, we will focus on the example
of decorated flippable plaquettes depicted in
Fig.~\ref{fig: decorated plaquette}.

The existence of the global expansion coefficients in the 
GS~(\ref{eq: GS general})
satisfying 
the integrability condition~(\ref{eq: integrability condition})
can be verified for the choice
\begin{equation}
\begin{split}
v(\mathbf{m}) 
&= 
\exp
\left(
u \delta N^{(f)}_{\ell^{*}_{0}}/2T
\right),
\\
\overline{v(\mathbf{m})} 
&= 
\exp
\left(
u \delta N^{(f)}_{\overline{\ell^{*}_{0}}}/2T
\right),
\\
t(\mathbf{m}) 
&= 
1, 
\end{split}
\end{equation}
with
\begin{equation}
\delta N^{(f)}_{\ell^{*}_{0}} 
= 
\pm [(m_1 + m_3) - (m_2 + m_4)] 
= 
- \delta N^{(f)}_{\overline{\ell^{*}_{0}}}.
\label{eq: deltaN def}
\end{equation}
Here, $u/T\in\mathbb{R}$ and
the $+$ ($-$) sign is associated to the vertical (horizontal)
orientation taken by the two parallel dimers occupying 
the flippable plaquette $p$ in $\ell^{*}_{0}$. One verifies that
$\delta N^{(f)}_{\ell^{*}_{0}}$ can only assume 
the values $0$, $\pm 1$, and $\pm 2$. 
With this choice, the Hamiltonian
\begin{equation}
\widehat{H}^{\ }_{0} 
:= 
\frac{1}{2} \sum^{\ }_{\ell^{*}_{0}}
\widehat{Q}^{\ }_{\ell^{*}_{0}}
\label{eq: def H0}
\end{equation}
is the sum over (non-commuting) operators with an index 
running over all decorated flippable plaquettes
\begin{equation}
\widehat{Q}^{\ }_{\ell^{*}_{0}}
:= 
    e^{u \delta N^{(f)}_{\ell^{*}_{0}}/2T} 
      \vert\ell^{*}_{0}\rangle\langle\ell^{*}_{0}\vert 
    + 
    e^{u \delta N^{(f)}_{\overline{\ell^{*}_{0}}}/2T} 
      \vert\overline{\ell^{*}_{0}}\rangle\langle\overline{\ell^{*}_{0}}\vert 
    - 
        \vert\overline{\ell^{*}_{0}}\rangle\langle\ell^{*}_{0}\vert 
    - 
        \vert\ell^{*}_{0}\rangle\langle\overline{\ell^{*}_{0}}\vert,
\label{eq: quantum Alet Hamiltonian}
\end{equation}
each of which is proportional to a projection operator: 
\begin{equation}
\widehat{Q}^{2}_{\ell^{*}_{0}} 
= 
\left(
e^{u \delta N^{(f)}_{\ell^{*}_{0}}/2T} 
+
e^{u \delta N^{(f)}_{\overline{\ell^{*}_{0}}}/2T} 
\right)
\widehat{Q}^{\ }_{\ell^{*}_{0}}.
\end{equation}
The nodeless wavefunction
\begin{equation}
\vert\Psi^{\ }_{0}\rangle 
= 
\sum_{\mathcal{C}\in\mathcal{S}^{\ }_{0}} 
  e^{\frac{u}{2T} N^{(f)}_{\mathcal{C}}} 
    \vert\mathcal{C}\rangle, 
\label{eq: quantum Alet GS}
\end{equation}
where $N^{(f)}_{\mathcal{C}}$ is the number of flippable plaquettes in the 
classical dimer configuration $\mathcal{C}$, 
is annihilated by the action of each $\widehat{Q}^{\ }_{\ell^{*}_{0}}$
for all $\ell^{*}_{0}$ as follows from verifying that
\begin{equation}
\delta N^{(f)}_{\ell^{*}_{0}} = 
N^{(f)}_{\overline{\mathcal{C}}}
-
N^{(f)}_{\mathcal{C}}.
\end{equation}
Therefore, $\vert\Psi^{\ }_{0}\rangle$ is a GS of
$\widehat{H}^{\ }_{0}$, which, with the help of
Perron-Fr{\"o}benius theorem, can be shown to be 
\emph{unique} within each \emph{irreducible} sector of 
$\widehat{H}^{\ }_{RK}$ under the action of 
$\widehat{H}^{\ }_{0}$~\cite{Castelnovo2005}. 
Remarkably, the GS expectation value 
of any quantum operator $\widehat{O}$ 
that is diagonal in the preferred basis $\{\vert\mathcal{C}\rangle\}$ 
can be written in term of an equilibrium 
thermal average for a square lattice classical dimer model,
\begin{equation}
\begin{split}
\frac{
\langle\Psi^{\ }_{0}\vert \widehat{O} \vert\Psi^{\ }_{0}\rangle 
     }
     {
\langle\Psi^{\ }_{0}\vert\Psi^{\ }_{0}\rangle 
     }
& = 
\sum_{\mathcal{C},\mathcal{C}'\in\mathcal{S}^{\ }_{0}} 
  e^{\frac{u}{2T} \Big( N^{(f)}_{\mathcal{C}}+N^{(f)}_{\mathcal{C}'} \Big)} 
\frac{
    \langle\mathcal{C}\vert \widehat{O} \vert\mathcal{C}'\rangle
     }
     {
\langle\Psi^{\ }_{0}\vert\Psi^{\ }_{0}\rangle 
     } 
\\
& = 
\frac{1}{Z(T/u)}
\sum_{\mathcal{C}\in\mathcal{S}^{\ }_{0}} 
  e^{\frac{u}{T} N^{(f)}_{\mathcal{C}}} 
    O^{\ }_\mathcal{C},
\end{split}
\label{eq: from quantum to classical expectation values}
\end{equation}
with
\begin{equation}
Z(T/u):=
\sum_{\mathcal{C}\in\mathcal{S}^{\ }_{0}} 
e^{-E^{(u)}_{\mathcal{C}}/T},
\qquad
E^{(u)}_{\mathcal{C}}:=-uN^{(f)}_{\mathcal{C}},
\label{eq: Alet partition function}
\end{equation}
and
\begin{equation}
O^{\ }_\mathcal{C}
:=
\langle\mathcal{C}\vert \widehat{O} \vert\mathcal{C}\rangle. 
\label{eq: def classical observable}
\end{equation}
It follows that the zero-temperature phase diagram of the 
interacting quantum SLDM (\ref{eq: def H0}) 
contains the phase diagram of the interacting classical 
SLDM (\ref{eq: Alet partition function}). 
The next section is devoted to the numerical study of the
phase diagram of the interacting classical 
SLDM~(\ref{eq: Alet partition function}). 
%
%

\section{\label{sec: classical model}
The associated classical model
           }
The interacting classical SLDM
defined by the partition function~(\ref{eq: Alet partition function}) 
has been extensively studied by Alet~\textit{et al.} 
in Ref.~\onlinecite{Alet2005} for positive 
values of the coupling constant $u$. Notice that in the range 
$K=u/T\in(0,\infty)$ the classical energy $E^{(u)}_{\mathcal{C}}$ favors 
configurations with a large number of flippable plaquettes, while the 
diagonal term in the quantum Hamiltonian in 
Eq.~(\ref{eq: def H0},\ref{eq: quantum Alet Hamiltonian}) 
always penalizes the presence of flippable plaquettes for all values of $K$! 
Conversely, in the range $K\in(-\infty,0)$ 
the presence of flippable plaquettes 
is also penalized at the classical level. 
Configurations with the largest number of flippable plaquettes are 
referred to as the \emph{columnar} state (every dimer has two parallel 
neighboring dimers along every other row or column). A representative
among all configurations with no flippable plaquettes is the
\emph{staggered} state. A staggered state is obtained 
from its parent columnar state upon translation of every other dimer 
of each column, say, by one lattice spacing along the direction 
parallel to the dimers. 
For brevity, we will refer to the parameter range $K>0$ as the 
\emph{columnar} side of the interaction, and to the parameter range $K<0$ as 
the \emph{staggered} side of the interaction. 

This section is devoted to studying the full parameter range 
$K\in(-\infty,\infty)$ 
of the interacting classical SLDM~(\ref{eq: Alet partition function}) 
using transfer matrix techniques. 
A brief summary of Alet's results is given in Sec.~\ref{sec: columnar side}, 
where we also present our results on the columnar side of the interaction. 
%
%

\subsection{\label{sec: CFT}
The conformal field theory description in the $T=\infty$ limit
           }
At infinite temperature ($K=0$), the partition 
function~(\ref{eq: Alet partition function}) 
reduces to the non-interacting classical SLDM. 
It exhibits critical spatial correlation functions 
with power-law decay%
~\cite{%
Kasteleyn1961,%
Fisher1961%
      }.
The long-wavelength limit of this model is known to be 
described by the $2D$ Sine-Gordon field theory
whose action can be written in terms of a continuous (height) scalar 
field $h$~\cite{Levitov1990,Kondev1995,Kondev1996},
\begin{equation}
S = \int d^{2}r 
  \left[
    \pi g | \nabla h(\mathbf{r}) |^2 
    + 
    V \cos\Big(2 \pi q \, h(\mathbf{r})\Big)
  \right], 
\label{eq: sine-Gordon action}
\end{equation}
where $g$ is called the stiffness and
the $V>0$ term is called the locking potential.
Remarkably, the stiffness and the periodicity of the locking
potential are fixed uniquely to the values $g=1/2$ and $q=4$,
respectively, if the Sine-Gordon action~(\ref{eq: sine-Gordon action})
is to encode the long distance asymptotics of the non-interacting
SLDM~\cite{Fisher1961,Kondev1995,Kondev1996}.
The scaling dimensions $d^{\ }_{e,m}$ 
of the so-called vertex operators at the free-field
fixed point $V=0$ of the Sine-Gordon
field theory~(\ref{eq: sine-Gordon action})
can be classified in terms of their \emph{electric} and \emph{magnetic} 
charge $e$ and $m$, respectively 
\begin{equation}
d^{\ }_{e,m} = \frac{1}{2} \left( \frac{e^2}{g} + g m^2 \right), 
\qquad\qquad 
e,m \in \mathbb{Z}. 
\label{eq: scaling dimensions}
\end{equation}
The physical interpretation of electric vertex operators is that their
correlation functions represent the long distance asymptotics of the
dimer correlation functions provided their charge $e$ is a multiple of $q=4$.
The physical interpretation of magnetic vertex operators is that their
correlation functions represent the long distance asymptotics of the
monomers correlation functions whereby it is understood that monomers
are defects in a dimer covering by which sites are not the end
points of dimers. Monomers will be introduced at the microscopic level
in Sec.~\ref{sec: monomers}
but are absent in the present SLDM, i.e., $m=0$ must be
enforced. If so, the most relevant electric
vertex operator with $e$ multiple of $q=4$ 
is the locking potential with scaling dimension
$d^{\ }_{4,0} = 16$ in the $K=0$ limit. We conclude that
the free-field fixed point $V=0$ is the attractive fixed-point of the 
Sine-Gordon theory~(\ref{eq: sine-Gordon action})
if it is to capture the long-distance physics of the non-interacting SLDM.
%
%

\subsection{\label{sec: transfer matrix}
Construction of the transfer matrix
           }
In order to study the phase diagram of the interacting classical 
SLDM~(\ref{eq: Alet partition function})
at finite values of the reduced coupling constant $K=u/T$,
it is convenient to use a combination of numerical transfer-matrix (TM) 
calculations in the \emph{infinite-strip geometry} 
with conformal field theory (CFT) arguments. 

\begin{figure}[ht]
\centering
\includegraphics[height=0.6\columnwidth]{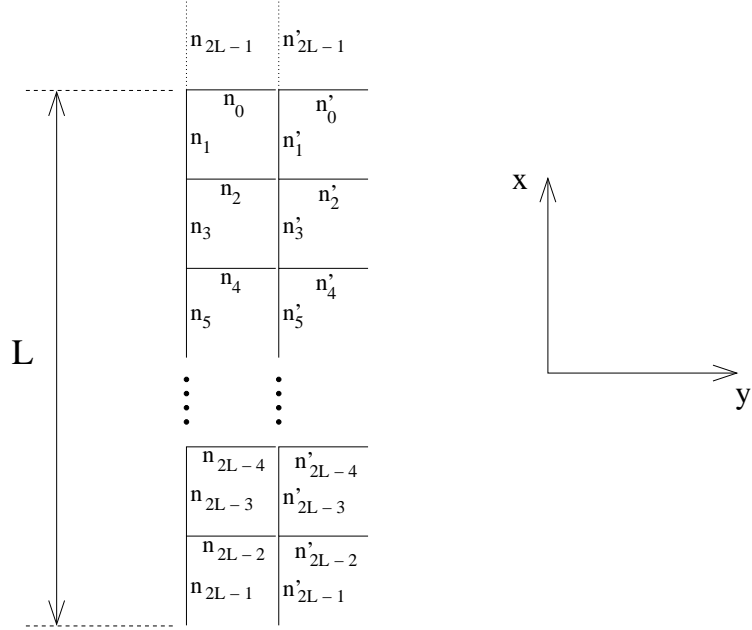}
\caption{
\label{fig: TM labeling}
Labeling of the edge variables $n^{\ }_i$ in the TM $T^{(L)}$. 
Periodic boundary conditions are assumed along the $x$ direction, 
i.e., $n^{\ }_0 \equiv n^{\ }_{2L}$.
}
\end{figure}

We define the interacting classical SLDM~(\ref{eq: Alet partition function})
on an $L\times M$, $M=\infty$ square lattice and impose
periodic boundary conditions in both directions (i.e., wrapped around a 
torus with infinite principal radius). The width $L$ must be \emph{even} to 
respect the bipartite nature of the lattice. The TM $T^{(L)}$ 
connects one row of the lattice to the following one along the 
principal (infinite) axis of the torus, as illustrated 
in Fig.~\ref{fig: TM labeling}, 
and satisfies
\begin{equation}
Z(T/u;L,M)=
\mathrm{Tr}\,
\left[T^{(L)}(T/u)\right]^{M}, 
\label{eq: def TM}
\end{equation} 
where $Z(T/u;L,M)$ is the partition function of the system. 
We shall assume (for simplicity) that the TM  $T^{(L)}$ can be diagonalized
through a similarity transformation, and we label its (positive) eigenvalues 
in descending order
\begin{equation}
T^{(L)}\sim
\mathrm{diag}\,
\left(
\begin{array}{cccc}
\Lambda^{(L)}_{0}
&
\Lambda^{(L)}_{1}
&
\Lambda^{(L)}_{2}
&
\cdots
\end{array}
\right),
\qquad
\Lambda^{(L)}_{0}
\geq
\Lambda^{(L)}_{1}
\geq
\Lambda^{(L)}_{2}
\geq
\cdots. 
\end{equation}
Anticipating an exponential growth with $L$ of the TM eigenvalues,
we also define the exponents
\begin{equation}
f^{\ }_{n}(L):=
-\frac{1}{L}
\ln\Lambda^{(L)}_{n},
\qquad
n=0,1,2,\cdots.
\end{equation}
The sign is chosen here by convention. 

The dimensionless intensive free energy in the thermodynamic limit
\begin{equation} 
f(T/u):=
\lim_{L\to\infty}f(T/u;L)
\end{equation}
is related to the TM $T^{(L)}$ by
\begin{equation}
f(T/u;L):= 
- 
\frac{1}{L} 
\lim_{M\to\infty} 
\frac{1}{M}
\ln
\textrm{Tr}
\left[
T^{(L)}(T/u)
\right]^{M}
\end{equation}
and is thus solely controlled by the largest (non-degenerate) eigenvalue
$\Lambda^{(L)}_{0}$ of the TM $T^{(L)}$, 
\begin{equation}
f(T/u)= 
-\lim_{L\to\infty}
\frac{1}{L}
\ln\,\Lambda^{(L)}_{0}(T/u)=
\lim_{L\to\infty}
f^{\ }_{0}(T/u;L).
\label{eq: TM free energy}
\end{equation}

If the long wavelength limit of the interacting classical 
SLDM~(\ref{eq: Alet partition function}) is captured by a CFT, the 
central charge of the CFT and the scaling dimensions of the primary fields 
in the CFT can also be extracted from the finite size ($L$) dependence 
of the TM $T^{(L)}$. The dependence on $T/u$ of the central charge $c(T/u)$ 
is given by
\begin{equation}
f(T/u;L)= 
f(T/u) 
- 
\frac{\pi c(T/u)}{6L^2} 
+
\mathcal{O}(1/L^{3}).
\label{eq: TM central charge}
\end{equation} 
The dependence on $T/u$ of the scaling dimensions $d^{\ }_{n}(T/u)$ 
of the CFT primary fields is in turn given by
\begin{equation}
f^{\ }_{n}(T/u;L)
- 
f(T/u;L)=
\frac{2\pi d^{\ }_{n}(T/u)}{L^2} 
+
\mathcal{O}(1/L^{3}),
\qquad
n=1,2,\cdots.
\label{eq: TM scaling dimension}
\end{equation}

We now turn to the explicit construction of the TM $T^{(L)}$.
To this end, we introduce the 
variables $n^{\ }_i$ on the bonds of the lattice,
$n^{\ }_i = 0\:(1)$ if the edge is 
empty (occupied) by a dimer, 
as is illustrated in Fig.~\ref{fig: TM labeling}. 
Any allowed (initial) configuration 
$\mathbf{n} = \{ n^{\ }_i,\;i=0,\, \ldots,\, 2L-1 \}$ 
must then satisfy
\begin{equation}
\label{eq: allowed conf condition}
n^{\ }_{2x-1} + n^{\ }_{2x} + n^{\ }_{2x+1} \leq 1, 
\qquad 
\forall\, x = 1,\, \ldots,\, L, 
\end{equation}
while the TM $T^{(L)}$
connects only configurations $\mathbf{n}$ and $\mathbf{n'}$ that 
satisfy
\begin{equation}
n'_{2x-1} + n'_{2x} + n'_{2x+1} + n^{\ }_{2x} = 1, 
\qquad 
\forall\, x = 1,\, \ldots,\, L.
\end{equation}
We can then write
\begin{equation}
T^{(L)}_{\mathbf{n}',\mathbf{n}}(T/u) 
= 
T^{(L)}_{\mathbf{n}',\mathbf{n}}(0) \;\; 
  U^{(L)}_{\mathbf{n}',\mathbf{n}}(T/u) \;\;  
    V^{(L)}_{\mathbf{n}',\mathbf{n}}(T/u), 
\label{eq: transfer matrix}
\end{equation}
where the two contributions due to the interaction 
$U^{(L)}_{\mathbf{n}',\mathbf{n}}(T/u)$ and 
$V^{(L)}_{\mathbf{n}',\mathbf{n}}(T/u)$ 
(accounting for horizontal and vertical parallel dimers, respectively), 
and the contribution due to the constraint 
$T^{(L)}_{\mathbf{n}',\mathbf{n}}(0)$ take the form
\begin{equation}
\begin{split}
T^{(L)}_{\mathbf{n}',\mathbf{n}}(0) 
& = 
\prod^{L}_{x=1} 
  \delta\left(n'_{2x-1} + n'_{2x} + n'_{2x+1} + n^{\ }_{2x} - 1\right), 
\\ 
U^{(L)}_{\mathbf{n}',\mathbf{n}}(T/u)  
& = 
\exp
  \left[ 
    \frac{u}{T} 
    \sum^{L-1}_{x=0} 
      \left(
        \frac{n^{\ }_{2x} n^{\ }_{2x+2} + n'_{2x} n'_{2x+2}}{2}
      \right) 
  \right],
\\ 
V^{(L)}_{\mathbf{n}',\mathbf{n}}(T/u) 
& = 
\exp
  \left[
    \frac{u}{T} 
    \sum^{L-1}_{x=0} n^{\ }_{2x+1} n'_{2x+1} 
  \right], 
\end{split}
\end{equation}
respectively.

The computational effort to obtain the eigenvalues of 
the TM $T^{(L)}$ can be reduced by looking for quantities that are 
left invariant under the action of the TM. 
This allows to block diagonalize 
$T^{(L)}(T/u)$ 
and to compute the largest eigenvalues separately in each sector. 
As discussed in Ref.~\onlinecite{Alet2005} and using the notation defined 
in Fig.~\ref{fig: TM labeling}, one can show that the quantity 
\begin{equation}
W^{\ }_{x} 
= 
\sum^{L/2-1}_{i=0} 
  \Big(
     n^{\ }_{4i} 
     - 
     n^{\ }_{4i+2} 
  \Big)
\;\; \in \;\; 
\left\{
  -\frac{L}{2},\, 
  \ldots,\, 
  \frac{L}{2} 
\right\}
\label{eq: def Wx}
\end{equation}
is conserved as one proceeds along the $y$-axis through repeated applications 
of the TM. One can thus use $W^{\ }_{x}$ to label the diagonal blocks 
of $T^{(L)}(T/u)$. Observe that any block with $W^{\ }_x \neq 0$ 
corresponds to having $|W^{\ }_x|$ monomers
on the same sublattice at $y=-\infty$, and $|W^{\ }_x|$ monomers on 
the opposite sublattice at $y=\infty$, as discussed in 
Ref.~\cite{Alet2005}. 
We will impose the condition that there are no monomers, i.e., $W^{\ }_x=0$, 
when considering the electrical sector of the CFT in isolation.
We shall assume that, upon ordering the eigenvalues of the TM 
within each sector according to 
\begin{equation}
\Lambda^{(W^{\ }_x,L)}_0 \geq \Lambda^{(W^{\ }_x,L)}_1 \geq \ldots,
\end{equation} 
$\Lambda^{(0,L)}_0$ is the global maximal TM eigenvalue. 
If so, it is $\Lambda^{(0,L)}_0$ that controls
the free energy and central charge of the system 
following Eq.~(\ref{eq: TM free energy}) 
and Eq.~(\ref{eq: TM central charge}) 
(with the possible exception of the 
transition to the staggered phase, as discussed 
in Sec.~\ref{sec: staggered side}). 
In Sec.~\ref{sec: uJ results summary} we show how the introduction of 
longer-range couplings in the dimer model gives rise to new phase transitions 
that are not entirely captured by the $W^{\ }_x=0$ sector, and one needs to 
compute the largest eigenvalue from the global TM. 
We shall also define the 
exponents 
\begin{equation}
f^{(W^{\ }_x)}_{n}(L):=
-
\frac{1}{L}\ln\Lambda^{(W^{\ }_x,L)}_{n},
\qquad
n=0,1,2,\cdots 
\end{equation}
within each irreducible sector. 
Equation~(\ref{eq: TM scaling dimension}) together with
Eq~(\ref{eq: scaling dimensions}) 
imply that the ratio 
$\Lambda^{(0,L)}_0/\Lambda^{(0,L)}_1$ 
determines the scaling dimension $d^{\ }_{1,0}$ 
while the ratio
$\Lambda^{(0,L)}_0/\Lambda^{(1,L)}_0$ 
determines the scaling dimension $d^{\ }_{0,1}$.
{}From the values of either of these two scaling dimensions one obtains the 
stiffness $g$ of the CFT of Eq.~(\ref{eq: sine-Gordon action}), and 
therefore the scaling dimensions of all the operators in the theory. 
Measuring both of them allows for a further check on the reliability of the 
numerical results, as the product of $d^{\ }_{0,1}$ and $d^{\ }_{1,0}$ 
must remain constant even if $g$ varies 
($d^{\ }_{0,1} \: d^{\ }_{1,0} = 1/4$). 
%
%

\subsection{\label{sec: results}
The transfer matrix results
           }
Exploiting the sparse nature of the TM due to the constraint-enforcing term 
$T^{(L)}_{\mathbf{n}',\mathbf{n}}(0)$, we compute its largest eigenvalues 
using a combination of hashing techniques to encode the state space of the 
system and routines from the ARPACK library~\cite{ARPACK} based on 
the implicitly restarted Arnoldi method (or implicitly restarted Lanczos 
method, whenever applicable), which are particularly suitable to handle 
large sparse matrices. 
Specifically, we compute the largest and first subleading eigenvalues of 
the TM in the $W^{\ }_x = 0$ sector, as well as the largest eigenvalue of 
the TM in the $W^{\ }_x = 1$ sector. 
This allows us to obtain, from finite size scaling, 
the dimensionless intensive free energies $f^{(0)}_{0}$ and $f^{(1)}_{0}$ 
in the sectors $W^{\ }_x = 0$ and $W^{\ }_x = 1$, respectively,  
the central charge $c$, 
and the scaling dimensions $d^{\ }_{1,0}$ and $d^{\ }_{0,1}$ of the 
electric and magnetic vertex operators, respectively. 
We also compute the global largest and first subleading eigenvalues of
the TM, i.e., independently of the value of $W^{\ }_x$.
We can thus verify that the global dimensionless intensive free energy
and the central charge are indeed controlled by $f^{(0)}_{0}$. 

For each value of the reduced coupling constant $K=u/T$ and in each 
block-diagonal sector, we consider systems of size 
$L = 6,\,8,\,10,\,12,\,14,\,16$. We then use standard finite-size 
scaling techniques~\cite{Jacobsen1998} to extrapolate the desired quantities. 
Namely, we fitted the computed values 
\begin{equation}
\begin{split}
f^{(0)}_{0}(L) 
& = 
- \frac{1}{L} \ln(\Lambda^{(0,L)}_0), 
\\ 
f^{(0)}_{1}(L) - f^{(0)}_{0}(L) 
& = 
- \frac{1}{L} \left[ \ln(\Lambda^{(0,L)}_1) 
                   - \ln(\Lambda^{(0,L)}_0) \right],
\\ 
f^{(1)}_{0}(L) - f^{(0)}_{0}(L) 
& = 
- \frac{1}{L} \left[ \ln(\Lambda^{(1,L)}_0) 
                   - \ln(\Lambda^{(0,L)}_0) \right], 
\end{split}
\label{eq: scaling forms from TM}
\end{equation}
for $L = 6,\,8,\,10,\,12,\,14,\,16$, with the scaling forms 
\begin{equation}
\begin{split}
f - \frac{\pi c}{6L^2} + \frac{A}{L^4}, 
\\
\frac{2\pi d_{1,0}}{L^2} + \frac{A}{L^4}, 
\\
\frac{2\pi d_{0,1}}{L^2} + \frac{A}{L^4}, 
\end{split}
\label{eq: scaling forms}
\end{equation}
respectively. The $1/L^4$ term is introduced to speed up the convergence 
of the fitting routine for the parameters $c$, $d_{1,0}$, and 
$d_{0,1}$~\cite{Jacobsen1998}. 
Specifically, we first obtain \emph{estimants} of the above parameters by 
fitting data points with 
$L^{\ }_{0} \leq L \leq L^{\ }_{\textrm{\small max}}$, 
where $L^{\ }_{\textrm{\small max}} = 16$ is the largest strip width that we 
consider, and $L^{\ }_{0}$ 
is varied from $6$ to $L^{\ }_{\textrm{\small max}}-4$ 
($L^{\ }_{\textrm{\small max}}-2$ in the case of the scaling exponents). 
The resulting estimants $c(L^{\ }_{0},L^{\ }_{\textrm{\small max}})$, 
$d_{1,0}(L^{\ }_{0},L^{\ }_{\textrm{\small max}})$, and 
$d_{0,1}(L^{\ }_{0},L^{\ }_{\textrm{\small max}})$ are 
given in Tables~\ref{table: c-estimants},~\ref{table: d10-estimants}, 
and~\ref{table: d01-estimants}. 
We then extrapolated the estimants in the limit 
$L^{\ }_{0} \to \infty$ by 
assuming the power law form~\cite{Jacobsen1998} 
\begin{equation}
x(L^{\ }_{0},L^{\ }_{\textrm{\small max}}) = x + k L^{-p}_0, 
\label{eq: pow law}
\end{equation}
for each of the three quantities $x=c,\,d_{1,0},\,d_{0,1}$, and performing 
an appropriate fit of the data in 
Tables~\ref{table: c-estimants},~\ref{table: d10-estimants}, 
and~\ref{table: d01-estimants}.
Whenever the last three estimants in a row of 
Tables~\ref{table: c-estimants},~\ref{table: d10-estimants}, 
and~\ref{table: d01-estimants} appear in monotonically decreasing 
order, they are used to obtain the constants $x$, $k$, and $p$. 
When this is not the case, or whenever the power $p$ obtained from the fit 
is too small to produce a reliable extrapolation 
(namely if $p<1$, from experience), then the Ising-like value $p=2$ is assumed 
and only the last two estimants are considered for the fit to obtain $x$ 
and $k$ (denoted by a $(\,)^*$ in the 
Tables~\ref{table: c-estimants},~\ref{table: d10-estimants}, 
and~\ref{table: d01-estimants})~\cite{Jacobsen1998}. 
A rough estimate for the error bar can be obtained by considering the 
variation among the estimants~\cite{Jacobsen1998}. 
In the results shown in 
Figs.~\ref{fig: c d columnar}-\ref{fig: free en staggered}, 
the error bars are computed 
as the square root of the variance of the last three estimants used for 
the extrapolation. In most cases, the error bars are smaller than the 
symbols used in the plots. 
\begin{table}
\caption{
\label{table: c-estimants}
List of the estimants $c(L^{\ }_{0},L^{\ }_{\textrm{\small max}})$
obtained from parabolic fits in $1/L^2$ for all the values of $K=u/T$ 
considered in this paper. The error bar on the extrapolated 
value is obtained from the variation among the estimants used in the 
corresponding extrapolation. 
}
\vspace{0.5 cm}
\centering
\begin{tabular}{cccccc}
\hline\hline\noalign{\smallskip}
    $K$ & $c(6,16)$ & $c(8,16)$ & $c(10,16)$ & $c(12,16)$ & Extrapolation \\ 
\noalign{\smallskip}\hline\noalign{\smallskip}
    1.7 & 0.9642 & 0.9510 & 0.9445 & 0.9380 & 0.923 $\pm$ 0.005$^*$ \\
    1.5 & 1.0135 & 1.0083 & 1.0099 & 1.0101 & 1.010 $\pm$ 0.001\\ 
    1.0 & 0.9900 & 0.9885 & 0.9924 & 0.9966 & 1.006 $\pm$ 0.003$^*$ \\ 
    0.7 & 0.9904 & 0.9886 & 0.9925 & 0.9966 & 1.006 $\pm$ 0.003$^*$ \\ 
    0.5 & 0.9906 & 0.9889 & 0.9923 & 0.9960 & 1.004 $\pm$ 0.003$^*$ \\ 
    0.4 & 0.9907 & 0.9891 & 0.9922 & 0.9955 & 1.003 $\pm$ 0.003$^*$ \\ 
    0.35 & 0.9910 & 0.9893 & 0.9934 & 0.9953 & 0.998 $\pm$ 0.002\\ 
    0.0 & 0.9947 & 0.9919 & 0.9951 & 0.9979 & 1.004 $\pm$ 0.002$^*$ \\ 
    -0.1 & 0.9959 & 0.9926 & 0.9950 & 0.9968 & 1.001 $\pm$ 0.002$^*$ \\ 
    -0.22 & 0.9974 & 0.9935 & 0.9944 & 0.9960 & 1.000 $\pm$ 0.001$^*$ \\ 
    -0.35 & 0.9985 & 0.9938 & 0.9952 & 0.9985 & 1.006 $\pm$ 0.002$^*$ \\ 
    -0.5 & 0.9991 & 0.9937 & 0.9951 & 0.9993 & 1.009 $\pm$ 0.002$^*$ \\ 
    -0.7 & 0.9963 & 0.9923 & 0.9932 & 0.9947 & 0.9981 $\pm$ 0.001$^*$ \\ 
    -1.0 & 0.9507 & 0.9716 & 0.9808 & 0.9864 & 1.005 $\pm$ 0.006 \\ 
    -1.12 & 0.8998 & 0.9401 & 0.9617 & 0.9741 & 1.007 $\pm$ 0.01 \\ 
    -1.28 & 0.7890 & 0.8563 & 0.8990 & 0.9276 & 0.99 $\pm$ 0.03$^*$ \\ 
    -1.5 & 0.5719 & 0.6472 & 0.7047 & 0.7491 & 0.85 $\pm$ 0.04$^*$ \\ 
    -2.0 & 0.2676 & 0.2765 & 0.2896 & 0.2991 & 0.321 $\pm$ 0.009$^*$ \\ 
    -3.0 & -0.0118 & -0.0086 & -0.0063 & -0.0068 & -0.008 $\pm$ 0.001 \\ 
    -4.0 & 0.0012 & -0.0006 & -0.0001 & -0.0001 & -0.0001 $\pm$ 0.0002\\ 
\noalign{\smallskip}\hline\hline
\end{tabular}
\end{table}
\begin{table}
\caption{
\label{table: d10-estimants}
List of the estimants $d_{1,0}(L^{\ }_{0},L^{\ }_{\textrm{\small max}})$ 
obtained from parabolic fits in $1/L^2$ for all the values of $K=u/T$ 
considered in this paper. The error bar on the extrapolated 
value is obtained from the variation among the estimants used in the 
corresponding extrapolation. 
}
\vspace{0.5 cm}
\centering
\begin{tabular}{cccccc}
\hline\hline\noalign{\smallskip}
    $K$ & $d_{1,0}(8,16)$ & $d_{1,0}(10,16)$ & $d_{1,0}(12,16)$ & 
          $d_{1,0}(14,16)$ & Extrapolation \\ 
\noalign{\smallskip}\hline\noalign{\smallskip}
    1.7 & 0.0878 & 0.0838 & 0.0806 & 0.0779 & 0.071 $\pm$ 0.002$^*$ \\ 
    1.5 & 0.1489 & 0.1483 & 0.1478 & 0.1475 & 0.1465 $\pm$ 0.0003$^*$ \\ 
    1.0 & 0.3095 & 0.3090 & 0.3087 & 0.3084 & 0.3076 $\pm$ 0.0003$^*$ \\ 
    0.7 & 0.4352 & 0.4350 & 0.4348 & 0.4347 & 0.4344 $\pm$ 0.0001$^*$ \\ 
    0.5 & 0.5454 & 0.5458 & 0.5461 & 0.5462 & 0.5467 $\pm$ 0.0002\\ 
    0.4 & 0.6120 & 0.6128 & 0.6132 & 0.6135 & 0.6141 $\pm$ 0.0003 \\ 
    0.35 & 0.6488 & 0.6498 & 0.6504 & 0.6507 & 0.6517 $\pm$ 0.0004 \\ 
    0.0 & 0.9986 & 0.9993 & 0.9995 & 0.9997 & 0.9998 $\pm$ 0.0002\\ 
    -0.1 & 1.1384 & 1.1381 & 1.1379 & 1.1379 & 1.1379 $\pm$ 0.0001\\ 
    -0.22 & 1.3375 & 1.3362 & 1.3359 & 1.3358 & 1.3358 $\pm$ 0.0001\\ 
    -0.35 & 1.5979 & 1.5998 & 1.6008 & 1.6012 & 1.6016 $\pm$ 0.0006\\ 
    -0.5 & 1.9848 & 1.9984 & 2.0018 & 2.0026 & 2.003 $\pm$ 0.002\\ 
    -0.7 & 2.1119 & 2.1336 & 2.1066 & 2.0768 & 1.99 $\pm$ 0.02$^*$ \\ 
    -1.0 & 1.9315 & 2.1446 & 2.3038 & 2.4152 & 3.0 $\pm$ 0.1\\ 
    -1.12 & 1.7082 & 1.9126 & 2.0810 & 2.2189 & 2.6 $\pm$ 0.1$^*$ \\ 
    -1.28 & 1.4078 & 1.5727 & 1.7151 & 1.8366 & 2.2 $\pm$ 0.1$^*$ \\ 
    -1.5 & 1.0517 & 1.1546 & 1.2495 & 1.3323 & 1.56 $\pm$ 0.07$^*$ \\ 
    -2.0 & 0.4538 & 0.4469 & 0.4471 & 0.4499 & 0.453 $\pm$ 0.001\\ 
    -3.0 & 0.9469 & 0.9481 & 0.9534 & 0.9605 & 0.980 $\pm$ 0.005$^*$ \\ 
    -4.0 & 1.6293 & 1.6312 & 1.6326 & 1.6340 & 1.638 $\pm$ 0.001$^*$ \\ 
\noalign{\smallskip}\hline\hline
\end{tabular}
\end{table}
\begin{table}
\caption{
\label{table: d01-estimants}
List of the estimants $d_{0,1}(L^{\ }_{0},L^{\ }_{\textrm{\small max}})$ 
obtained from parabolic fits in $1/L^2$ for all the values of $K=u/T$ 
considered in this paper. The error bar on the extrapolated 
value is obtained from the variation among the estimants used in the 
corresponding extrapolation. 
}
\vspace{0.5 cm}
\centering
\begin{tabular}{cccccc}
\hline\hline\noalign{\smallskip}
    $K$ & $d_{0,1}(8,16)$ & $d_{0,1}(10,16)$ & $d_{0,1}(12,16)$ & 
          $d_{0,1}(14,16)$ & Extrapolation \\ 
\noalign{\smallskip}\hline\noalign{\smallskip}
    1.7 & 1.8139 & 1.9017 & 1.9671 & 2.0186 & 2.16 $\pm$ 0.05$^*$ \\ 
    1.5 & 1.4269 & 1.4689 & 1.4972 & 1.5175 & 1.57 $\pm$ 0.02$^*$ \\ 
    1.0 & 0.8044 & 0.8095 & 0.8121 & 0.8136 & 0.817 $\pm$ 0.002 \\ 
    0.7 & 0.5735 & 0.5747 & 0.5752 & 0.5755 & 0.5759 $\pm$ 0.0003 \\ 
    0.5 & 0.4562 & 0.4567 & 0.4569 & 0.4570 & 0.4572 $\pm$ 0.0001 \\ 
    0.4 & 0.4061 & 0.4064 & 0.4066 & 0.4067 & 0.4067 $\pm$ 0.0001 \\ 
    0.35 & 0.3829 & 0.3832 & 0.3833 & 0.3834 & 0.3835 $\pm$ 0.0001 \\ 
    0.0 & 0.2494 & 0.2497 & 0.2498 & 0.2498 & 0.2499 $\pm$ 0.0001 \\ 
    -0.1 & 0.2191 & 0.2194 & 0.2195 & 0.2196 & 0.2196 $\pm$ 0.0001 \\ 
    -0.22 & 0.1865 & 0.1868 & 0.1869 & 0.1870 & 0.1871 $\pm$ 0.0001 \\ 
    -0.35 & 0.1554 & 0.1557 & 0.1558 & 0.1559 & 0.1559 $\pm$ 0.0001 \\ 
    -0.5 & 0.1243 & 0.1246 & 0.1248 & 0.1248 & 0.1249 $\pm$ 0.0001 \\ 
    -0.7 & 0.0901 & 0.0902 & 0.0904 & 0.0905 & 0.0908 $\pm$ 0.0001 \\ 
    -1.0 & 0.05186 & 0.05166 & 0.05164 & 0.05164 & 0.05164 $\pm$ 0.00001 \\ 
    -1.12 & 0.0404 & 0.0400 & 0.0399 & 0.0399 & 0.0398 $\pm$ 0.0001 \\ 
    -1.28 & 0.0279 & 0.0275 & 0.0273 & 0.0271 & 0.0266 $\pm$ 0.0002$^*$ \\ 
    -1.5 & 0.0089 & 0.0110 & 0.0147 & 0.0145 & 0.014 $\pm$ 0.002$^*$ \\ 
    -2.0 & -0.0099 & -0.0005 & -0.0052 & -0.0042 & -0.001 $\pm$ 0.002 \\ 
    -3.0 & -0.0011 & -0.0003 & 0.0002 & 0.0002 & 0.0001 $\pm$ 0.0003 \\ 
    -4.0 & 0.00008 & 0.00002 & 0.00000 & 0.00000 & 0.00000 $\pm$ 0.00001 \\ 
\noalign{\smallskip}\hline\hline
\end{tabular}
\end{table}

The results for the central charge and scaling exponents are shown in 
Figs.~\ref{fig: c d columnar} and~\ref{fig: c d staggered}, for the 
columnar and staggered side of the interaction respectively. 

The same fitting approach is used to extrapolate the global free energy 
$f$ from its estimants $f(L^{\ }_{0},L^{\ }_{\textrm{\small max}})$, as well as 
$f^{(0)}_{\ }$ (in the sector $W^{\ }_{x}=0$) and 
$f^{(1)_{\ }}$ (in the sector $W^{\ }_{x}=1$). 
The data are omitted here for brevity, and we show only the 
results for the staggered side in Fig.~\ref{fig: free en staggered}. 
%
%

\subsubsection{\label{sec: columnar side}
The columnar side
           }
\begin{figure}[ht]
\centering
\includegraphics[width=0.6\columnwidth]{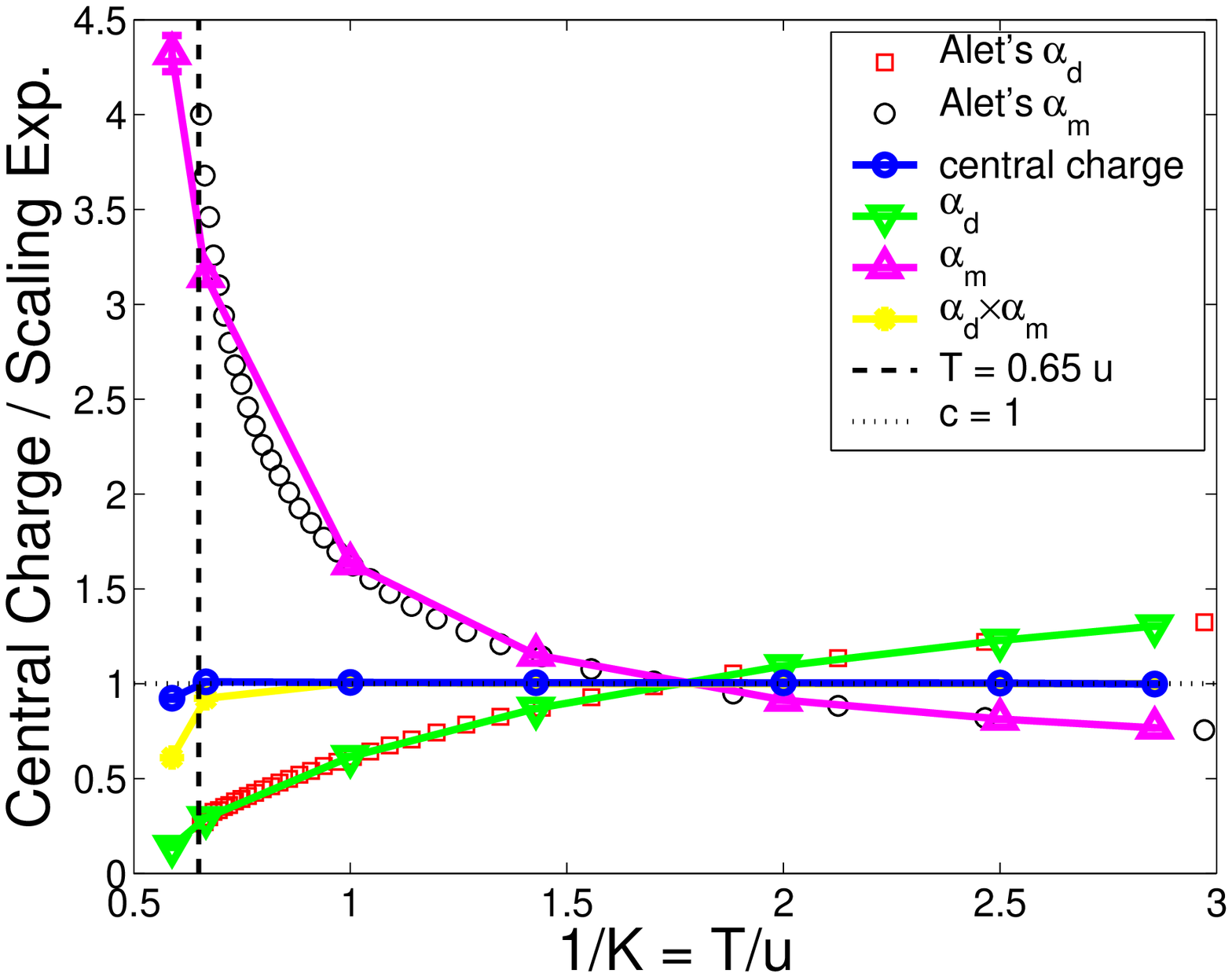}
\caption{
\label{fig: c d columnar}
Central charge and scaling exponents for the electric and magnetic 
monopole operators on the columnar side of the interaction. 
Following the convention in Ref.~\onlinecite{Alet2005}, 
we plot the scaling exponents $\alpha^{\ }_d = 2\,d^{\ }_{1,0}$ and 
$\alpha^{\ }_m = 2\,d^{\ }_{0,1}$ (equal to $1/g$ and $g$ from 
Eq.~(\ref{eq: scaling dimensions}), respectively), 
instead of the scaling dimensions $d^{\ }_{e,m}$. 
The results by Alet \textit{et al.} are also shown for 
comparison~\cite{Alet2005}. 
}
\end{figure}

On the columnar side of the interaction (Fig.~\ref{fig: c d columnar}), 
our results are in good agreement with the ones obtained by 
Alet \textit{et al.}. 
The system remains critical for small but finite 
values of the reduced coupling $K=u/T$ up to some critical value 
in the range $(1.5,1.7)$ (compare with the critical temperature 
$T^{\textrm{\scriptsize (columnar)}}_c \simeq 0.65\,u$ in 
Ref.~\onlinecite{Alet2005}) and its long 
wavelength limit is described by the CFT%
~(\ref{eq: sine-Gordon action}) with continuously varying stiffness 
$g \equiv 2\,d_{0,1}$. 
In the limit of $K\to0$, one recovers the expected value $g=1/2$ (not shown)
and the cosine term is irrelevant as $d_{4,0}=8/g=16$. 
When the value of $K$ is increased, the stiffness increases monotonically 
and the scaling dimension of the cosine term decreases correspondingly, 
until this operator becomes marginal at 
$T=T^{\textrm{\scriptsize (columnar)}}_c$ ($d_{4,0}=2$, $g=4$). The 
system undergoes then a Kosterlitz-Thouless (KT) transition into the 
columnar ordered phase, as confirmed by the Monte Carlo simulation results 
presented in Ref.~\onlinecite{Alet2005}. 

Notice that, contrary to the RK model in 
Eq.~(\ref{eq: RK Hamiltonian})~\cite{Syljuasen2005}, 
in our quantum model the presence of a possible resonating plaquette phase 
seems to be excluded according to the Monte Carlo results by 
Alet~\textit{et al.}~\cite{Alet2005}. 
%
%

\subsubsection{\label{sec: staggered side}
The staggered side
           }
\begin{figure}[ht]
\centering
\includegraphics[width=0.6\columnwidth]{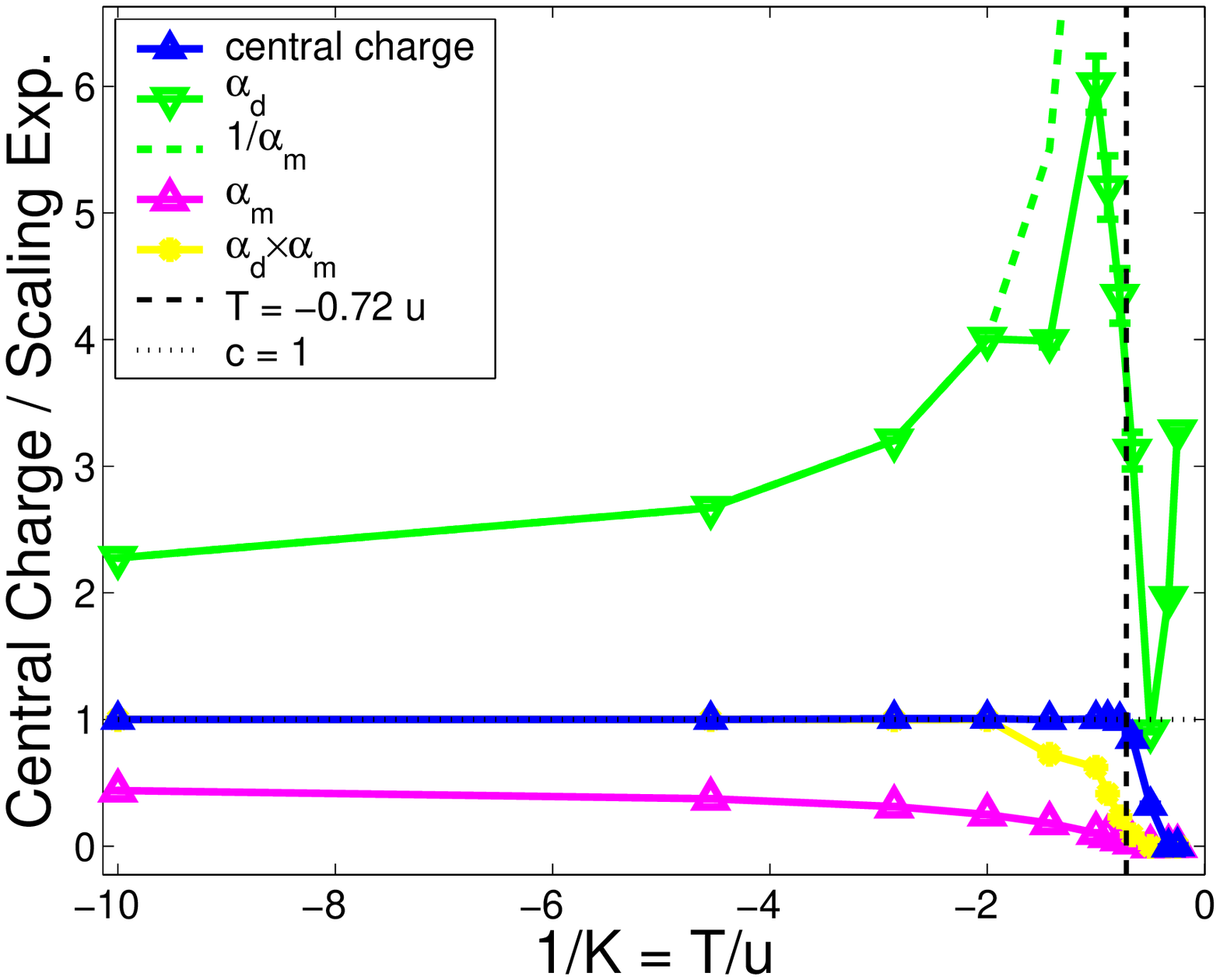}
\caption{
\label{fig: c d staggered}
Central charge and scaling exponents for the electric and magnetic 
monopole operators on the staggered side of the interaction. 
Following the convention used in Ref.~\onlinecite{Alet2005}, 
we plot the scaling exponents $\alpha^{\ }_d = 2\,d^{\ }_{1,0}$ and 
$\alpha^{\ }_m = 2\,d^{\ }_{0,1}$ (equal to $1/g$ and $g$ from 
Eq.~(\ref{eq: scaling dimensions}), respectively), 
instead of the scaling dimensions $d^{\ }_{e,m}$. 
The divergence of $\alpha^{\ }_d$ close to 
$T/u\simeq-1$ makes it increasingly 
difficult to obtain a reliable extrapolation for $L\to\infty$, 
hence the large fluctuations observed. 
The smaller parameter $\alpha_m$ is much less affected and the curve 
$1/\alpha_m$ can be used as a guide to the eye for the diverging behavior 
of $\alpha_d$. 
}
\end{figure}

On the staggered side of the interaction (Fig.~\ref{fig: c d staggered}), 
the central charge remains again constant at $c=1$ for small but finite 
values of the reduced coupling $K$, until it abruptly drops to zero at 
some critical value in the range $(-1.5,-1.28)$. 
We can therefore estimate the critical temperature for the transition to 
the staggered phase to be 
$T=T^{\textrm{\scriptsize (staggered)}}_c \simeq (-0.72 \pm 0.05)\,u$. 
The system remains critical for temperatures above 
$T^{\textrm{\scriptsize (staggered)}}_c$ while the stiffness $g$ decreases 
monotonically with decreasing $T$, as signaled by the behavior of the 
scaling dimensions $d_{1,0}$ and $d_{0,1}$. 

Notice that, on this side of the interaction, all the allowed operators with 
zero magnetic charge (i.e., those with $e = 4n$, $n = 1,2,\ldots$) 
have scaling dimensions $d_{e,0}$ larger than two. 
The absence of an allowed operator that becomes relevant for 
$T>T^{\textrm{\scriptsize (staggered)}}_c$ is consistent with the transition 
being first order, as expected in analogy with the RK Hamiltonian%
~\cite{Rokhsar1988} (see also the results on the free energy presented below). 

\begin{figure}[ht]
\centering
\includegraphics[width=0.6\columnwidth]{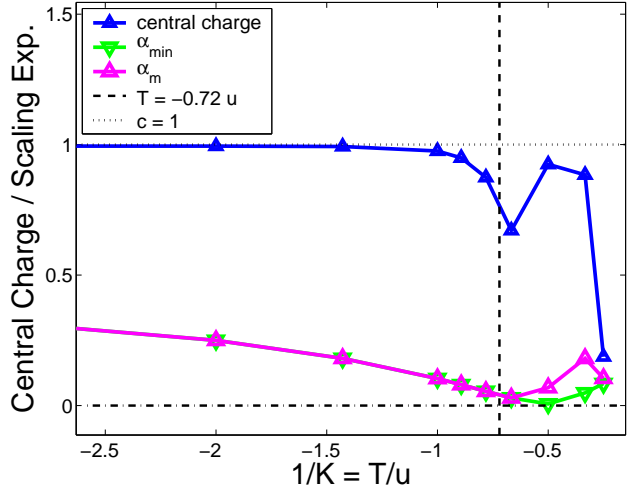}
\caption{
\label{fig: c d staggered dg}
Estimants of the central charge and scaling exponent for the 
magnetic monopole operator on the staggered side of the interaction, 
derived using the global transfer matrix for 
$L=8,12,16$~\cite{note: staggered caveat}. 
Following the convention used in Ref.~\onlinecite{Alet2005}, 
we plot the scaling exponents $\alpha^{\ }_d = 2\,d^{\ }_{1,0}$ and 
$\alpha^{\ }_m = 2\,d^{\ }_{0,1}$ (equal to $1/g$ and $g$ from 
Eq.~(\ref{eq: scaling dimensions}), instead of the scaling dimensions 
$d^{\ }_{e,m}$. 
For comparison, we also show the estimants of 
$\alpha^{\ }_{\textrm{min}} = 2\,d^{\ }_{\textrm{min}}$ corresponding to the 
most relevant operator in the CFT, obtained from the first subleading 
eigenvalue of the global TM. 
As expected, $\alpha^{\ }_m = \alpha^{\ }_{\textrm{min}}$ as long as the CG 
description in Sec.~\ref{sec: CFT} holds, i.e., for 
$|T/u|>|(T/u)^{\textrm{\scriptsize (staggered)}}_c|$. 
}
\end{figure}

Finally, the stiffness $g$ seems to vanish exactly at the staggered 
transition (within numerical error), suggesting that a tilting transition 
occurs at $T^{\textrm{\scriptsize (staggered)}}_c$ as briefly noted in 
Ref.~\onlinecite{Alet2005}. Observe, however, that our estimate 
$
T^{\textrm{\scriptsize (staggered)}}_c 
\simeq 
(-0.72 \pm 0.05) u
$
differs from the estimate
$
T^{\textrm{\scriptsize (staggered)}}_c 
\simeq 
-0.449(1) u
$
in Ref.~\onlinecite{Alet2005}.
A possible explanation for this discrepancy might be due to strong
topology-dependent finite-size effects. 
For example, the central charge of the system computed
using the global transfer matrix as is done in Fig.~\ref{fig: c d staggered dg}
suggests a drop of the central charge at the much lower value of $-0.3u$
than the value $-0.72u$ extracted from Fig.~\ref{fig: c d staggered}.
In turn, this suggests a strong dependency on the topological sectors
of the finite-size estimate for the critical temperature.
Understanding this interplay between finite-size effects and
topological sectors is beyond the scope of this paper.
%
%

\subsubsection{\label{sec: free energy}
The free energy
           }
\begin{figure}[ht]
\centering
\includegraphics[width=0.6\columnwidth]{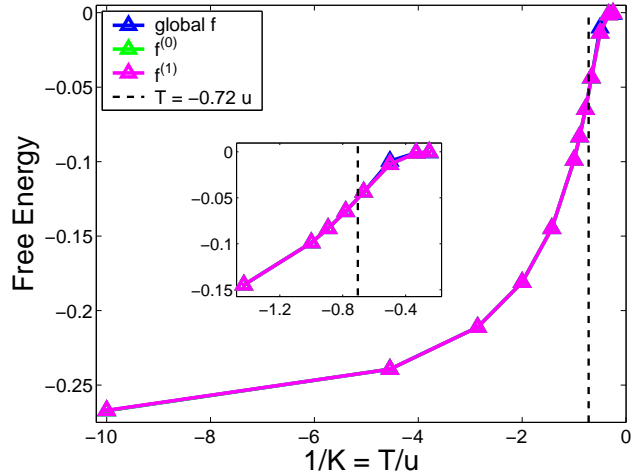}
\caption{
\label{fig: free en staggered}
Free energy of the system obtained from the global TM and from the 
TM restricted to the $W^{\ }_x=0$ and $W^{\ }_x=1$ sectors, respectively, 
for the staggered side of the interaction. The inset shows a close-up 
view of the temperature range where the transition is expected to happen, 
and the dashed line indicates the value of 
$T^{\textrm{\scriptsize (staggered)}}_c$ predicted from CFT arguments in 
Sec.~\ref{sec: staggered side}. 
}
\end{figure}

The results for the global free energy of the system ($f$) and for the 
constrained free energies in the $W^{\ }_x=0$ ($f^{(0)}_{0}$) and 
$W^{\ }_x=1$ ($f^{(1)}_{0}$) sectors are shown in 
Fig.~\ref{fig: free en staggered} for the staggered side of the 
interaction. 
The continuous KT transition on the columnar side is in general difficult to 
detect from free energy measurements, and the curves (not shown) are indeed 
smooth and featureless across $T^{\textrm{\scriptsize (columnar)}}_c$. 

On both sides of the interaction and for the range of temperatures 
considered here, the three free energies are indistinguishable 
($f=f^{(0)}_{0}=f^{(1)}_{0}$). 
This overlap is indeed expected since the number of $W^{\ }_x$ sectors is 
linear in system size, and the presence of defects at $y=\pm \infty$ should 
not affect significantly the entropy of each sector. 

As mentioned above, the transition on the staggered side is expected to be 
first order. In fact, not only has the set of all possible staggered 
configurations vanishing entropy, but they also do not allow for local 
fluctuations (without violating the dimer constraint), so that even the set 
of all the configurations connected to a staggered one through a single thermal 
fluctuation has vanishing entropy. 
This peculiar connectivity of phase space close to the staggered phase leaves 
little room for a possible continuous phase transition, and indeed in 
the RK Hamiltonian~(\ref{eq: RK Hamiltonian}) the 
transition to the staggered phase is similarly observed to be first order. 

The accuracy in our data does not allow for a precise identification of 
the transition temperature from the free energy plot. However, the overall 
behavior is in agreement with the value of 
$T^{\textrm{\scriptsize (staggered)}}_c$ obtained from CFT considerations, 
as shown in the inset in Fig.~\ref{fig: free en staggered} 
(notice that the staggered phase has both vanishing entropy and vanishing 
energy, therefore a lower bound for the transition temperature is obtained 
from the location of the point where the free energy first vanishes). 
A deeper insight on the nature of this transition could perhaps be obtained 
from Monte Carlo simulations, even though the non-local nature of the thermal 
fluctuations above the staggered phase are likely to lead to dynamical 
slowing down and glassiness, as observed in similar 
coloring models~\cite{Castelnovo2004}. 
A variational (cluster) mean field approach~\cite{Cirillo1996,Castelnovo2004} 
may prove more powerful in obtaining an accurate estimate of the transition 
temperature to compare with the one from CFT arguments. 
Such analysis is however beyond the scope of the present paper. 

An intriguing question about this transition to the staggered phase would be 
to understand what happens in this model to the so-called devil's staircase 
scenario, proposed in the original Rokhsar-Kivelson square 
lattice dimer model~(\ref{eq: RK Hamiltonian})~\cite{Fradkin2004}.
%
%

\section{\label{sec: long range interactions}
Stability of the critical line
        }
\begin{figure}[t]
\centering
\includegraphics[width=0.6\columnwidth]{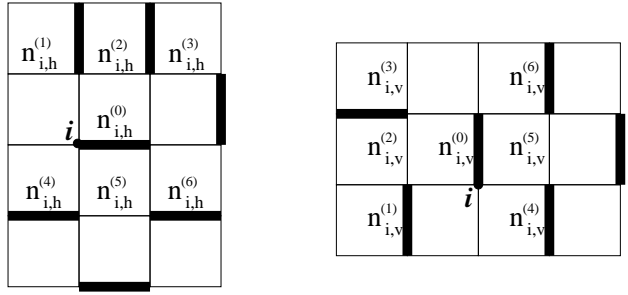}
\caption{
\label{fig: nnn term labeling}
Given any site $i$ of the square lattice, 
the locations of the horizontal bond variables
$\mathbf{n}^{\ }_{i,h}:=\left(n^{(0)}_{i,h},\cdots,n^{(6)}_{i,h}\right)$
and the locations of the vertical bond variables
$\mathbf{n}^{\ }_{i,v}:=\left(n^{(0)}_{i,v},\cdots,n^{(6)}_{i,v}\right)$
are defined in the left and right panel, respectively.
Observe that the right panel is obtained from the left panel
after a $\pi/2$ counterclockwise rotation followed by the
substitution $h\to v$. Given the lattice site $i$,
$r\in\{0,\cdots,6\}$,
and
$\alpha\in\{h,v\}$ 
the bond variable
$n^{(r)}_{i,\alpha}$
takes the value $1$ if a horizontal ($\alpha=h$) 
or vertical ($\alpha=v$) dimer is present
while $n^{(r)}_{i,\alpha}=0$ otherwise.
The contribution from site $i$ to the classical energy 
$E^{(J)}_{\mathcal{C}}$ defined in Eq.~(\ref{eq: J interaction})
is
$-J$ for the dimer covering $\mathcal{C}$ in the left panel and
$-3J/2$ for the dimer covering $\mathcal{C}$ in the right panel.
        }
\end{figure}

\begin{figure}[ht]
\centering
\includegraphics[width=0.75\columnwidth]{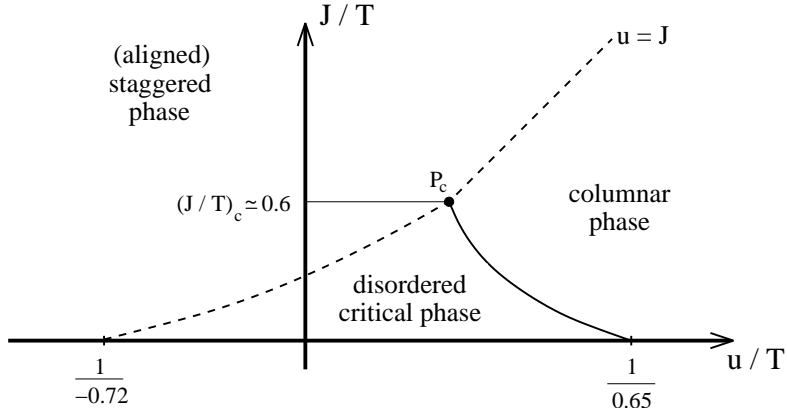}
\caption{
\label{fig: uJ phase diag}
Phase diagram of the dimer model with both plaquette counting ($u$) and 
short-range aligning ($J$) interactions. The dashed line delimiting the 
staggered phase is expected to be a coexistence line (a first-order phase 
transition), while the solid line between the columnar and the disordered 
critical phase is expected to be a line of KT (infinite order) transition 
points. 
The quantitative details shown here are derived from numerical and 
analytical arguments in Sec.~\ref{sec: uJ results summary}. 
Notice that the phase diagram of a classical system with fixed values of 
$u$ and $J$ as a function of temperature is described by a straight line 
that passes by the origin of the axes. 
        }
\end{figure}

The question that we want to address in this section is how stable
is the critical segment $-1/0.72\lesssim u/T \lesssim 1/0.65$
in the zero-temperature phase diagram of
the interacting quantum SLDM~(\ref{eq: def H0})
to a perturbation by some longer-range interaction $E^{(J)}_{\mathcal{C}}$ 
between dimers. 
In order to address this question, at least from a qualitative point of 
view, we need to substitute 
\begin{equation}
E^{(u)}_{\mathcal{C}}\longrightarrow
E^{(u)}_{\mathcal{C}}
+
E^{(J)}_{\mathcal{C}}
\end{equation}
in the classical partition function~(\ref{eq: Alet partition function})
by retracing all the steps that lead from the interacting quantum 
SLDM~(\ref{eq: def H0})
to the classical partition function~(\ref{eq: Alet partition function}).
(That this is possible was proven in Ref.~\cite{Castelnovo2005}.) 
For the sake of concreteness, let us consider the case where 
the new contribution to the classical energy of a dimer configuration
$\mathcal{C}$ is given by 
\begin{equation}
E^{(J)}_{\mathcal{C}}= 
-\frac{J}{2}  \sum_{i}\sum_{\alpha=h,v} n^{(0)}_{i \alpha} 
  \left[
   n^{(1)}_{i \alpha} + 
   n^{(2)}_{i \alpha} + 
   n^{(3)}_{i \alpha} + 
   n^{(4)}_{i \alpha} + 
   n^{(5)}_{i \alpha} + 
   n^{(6)}_{i \alpha} 
  \right]. 
\label{eq: J interaction}
\end{equation}
Here, the sum over $i$ runs over all lattice sites,
the sum over $\alpha=h,v$ runs over the occupation number by
horizontal and vertical dimers respectively, and
$n^{(r)}_{i \alpha} = 1\:(0)$, $r=0,\cdots,6$ if the bonds depicted
in Fig.~\ref{fig: nnn term labeling} are covered (not covered) by a dimer.  
Notice that the $J$ interaction is comprised of a nearest-neighbor part 
$$
n^{(0)}_{i \alpha} n^{(2)}_{i \alpha} 
+ 
n^{(0)}_{i \alpha} n^{(5)}_{i \alpha}
$$ 
and of a next-nearest-neighbor part 
$$
n^{(0)}_{i \alpha} n^{(1)}_{i \alpha}
+ 
n^{(0)}_{i \alpha} n^{(3)}_{i \alpha}
+ 
n^{(0)}_{i \alpha} n^{(4)}_{i \alpha} 
+ 
n^{(0)}_{i \alpha} n^{(6)}_{i \alpha}
$$ 
that are equally weighted. 
Therefore, varying both $u$ -- which corresponds to a 
pure nearest-neighbor interaction between dimers -- and $J$ is equivalent to 
tuning separately the nearest-neighbor and next-nearest-neighbor couplings 
between dimers. 
In the limit $J/T \gg |u|/T$, the coupling $J$ favors dimer configurations 
where the dimers are mostly aligned along the same direction.
In this limit, the disordered critical phase between the two 
ordered phases is penalized. For this reason
we conjecture that the effect of an increasing value of $J/T>0$ is to
shrink continuously the size of the critical segment 
$-1/0.72\lesssim u/T\lesssim 1/0.65$ at $J=0$ 
until the critical value $(J/T)^{\ }_{\mathrm{c}}$ is reached
at which the critical line terminates into a tricritical point.
A similar behavior is observed in the classical phase diagram of the 
three-coloring model in presence of an Ising interaction and of a 
uniform magnetic field~\cite{Cirillo1996}.
This scenario is depicted 
in Fig.~\ref{fig: uJ phase diag}. 
Beyond this point, i.e.,
for $J/T>(J/T)^{\ }_{\mathrm{c}}$,
we conjecture that the columnar and staggered phases are separated by
a first-order phase transition. 
The slope of the line of first-order transition points  
for  $J/T>(J/T)_{\mathrm{c}}$ is positive because the 
$J>0$ coupling favors the staggered phase over the columnar phase. 
The point $P^{\ }_{\mathrm{c}}$, 
where the continuous KT phase transition to the columnar phase
meets the first order phase transition to the staggered phase, 
is reminiscent of the RK critical point~(\ref{eq: def RK point}) 
in the RK Hamiltonian~(\ref{eq: RK Hamiltonian}) in that
the transition is first order on one side and continuous on the other 
side~\cite{note: RK resonating}.

These conjectures have been tested numerically with the help
of transfer matrix calculations of the central charge and 
scaling exponents for the system in presence of the additional energy term 
$E^{(J)}_{\mathcal{C}}$ 
in 
Eq.~(\ref{eq: J interaction}). 
The numerical techniques that we shall use follow step by step those described in 
Sec.~\ref{sec: classical model}, 
provided we introduce two extra factors 
$\tilde{U}^{(L)}_{\mathbf{n}',\mathbf{n}}(T/J)$ and 
$\tilde{V}^{(L)}_{\mathbf{n}',\mathbf{n}}(T/J)$ 
in Eq.~(\ref{eq: transfer matrix}), 
\begin{equation}
\begin{split}
\tilde{U}^{(L)}_{\mathbf{n}',\mathbf{n}}(T/J)  
& = 
\exp
  \left[ 
    \frac{J}{T} 
    \sum^{L-1}_{x=0} 
      \frac{n^{\ }_{2x}}{2} \times 
\right.
\\
& \qquad 
      \left( \vphantom{\Big(}
        (1 - n^{\ }_{2x-1} + n^{\ }_{2x-2} + n^{\ }_{2x-3}) 
	+ 
        n^{\ }_{2x-2} + n^{\prime}_{2x-2}
\right.
\\ 
& \qquad
\left.
\left.  +
	(1 - n^{\ }_{2x+1} + n^{\ }_{2x+2} + n^{\ }_{2x+3}) 
	+ 
	n^{\ }_{2x+2} + n^{\prime}_{2x+2}
      \vphantom{\Big(} \right) 
\vphantom{\sum^{L-1}_{x=0}}
  \right],
\\ 
\tilde{V}^{(L)}_{\mathbf{n}',\mathbf{n}}(T/J)  
& = 
\exp
  \left[ 
    \frac{J}{T}
    \sum^{L-1}_{x=0} 
      n^{\ }_{2x+1} 
      \left( \vphantom{\Big(}
        n^{\prime}_{2x-1} + n^{\prime}_{2x+1} + n^{\prime}_{2x+3} 
      \right) 
  \vphantom{\sum^{L-1}_{x=0}} \right],
\end{split}
\end{equation}
for horizontal and vertical dimers respectively. 
The notation used here for the labeling of the bonds in the lattice is the 
same as in Sec.~\ref{sec: transfer matrix} 
(see Fig.~\ref{fig: TM labeling}). 
The results obtained from these numerical calculations are summarized in
Sec.~\ref{sec: uJ results summary} and are in agreement with the conjectured phase 
diagram in Fig.~\ref{fig: uJ phase diag}. 
%
%

\subsection{\label{sec: uJ results summary}
Numerical results
           }
We ran transfer matrix simulations at fixed values of 
$J/T = -0.2,\, 0.2,\, 0.5,\, 1.0$, and $1.5$, 
while varying $u/T$ along the same set of values used for the $J=0$ case. 
We also sampled the line $u=J(=1)$ in parameter space, at different 
(inverse) temperatures 
$
1/T 
= 
0,\, 0.1,\, 0.2,\, 0.3,\, 0.4,\, 0.5,\, 0.7,\, 
1.0,\, 1.2,\, 1.5,\, 1.7,\, 2.0,\, 2.5,\, 3.0
$. 

Before presenting the numerical results, let us discuss the low-temperature 
asymptotics of the system for which analytical arguments can be formulated. 
For the sake of simplicity we will discuss here the case of an $L \times L$
square lattice, which one can then generalize to the $M \times L$ case 
appropriate for TM calculations. As before, $L$ is chosen even when
imposing periodic boundary conditions so as to maintain 
bipartiteness of the square lattice.

\begin{figure}[ht]
\centering
\subfigure[\label{fig: staggered-Ising}
Staggered phase degeneracy for $J=0$
          ]
{\includegraphics[width=0.7\columnwidth]{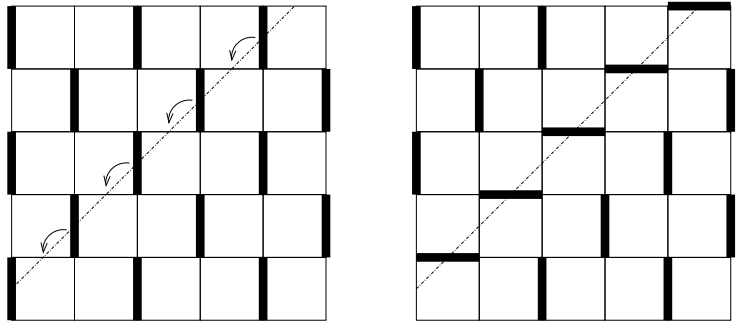}}
\\ \vspace{0.0 cm}
\subfigure[\label{fig: columnar-Ising}
Columnar phase degeneracy for $u=J$
          ]
{\includegraphics[width=0.7\columnwidth]{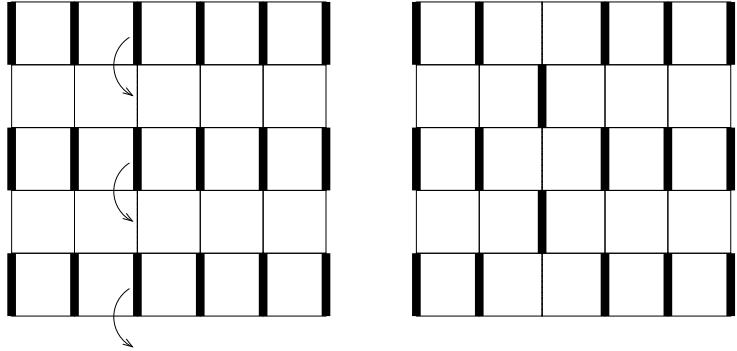}}
\caption{
\label{fig: asymptotic phases}
Examples of ordered phases in the asymptotic limits $u/T \to -\infty, J=0$ 
(Top) and $u/T, J/T \to \infty$ (Bottom). 
For $J=0$ all staggered configurations are degenerate, and can be obtained 
from the four canonical configurations -- one of which is shown in the 
Top-left figure -- by $\pi/2$-degree rotation of the dimers along diagonals 
(Top-right). 
Equivalently, for $u = J$ all configurations obtained from one 
of the four columnar states by one-sublattice translations of rows of 
collinear dimers (Bottom-left $\leftrightarrow$ Bottom-right) are also 
degenerate. 
        }
\end{figure}

We begin by considering the case of $u/T\ll -1$ and $J/T=0$ 
when the system lies deep within the staggered phase. 
The degeneracy of the classical ground state manifold 
in the staggered phase is \emph{subextensive}, in the sense that the number 
of degenerate configurations is exponential in the \emph{linear} size 
of the system.
To see this start from the staggered configuration
defined in the left panel of Fig.~\ref{fig: staggered-Ising}.
Draw a line that crosses the mid-points of the occupied bonds along a 
diagonal, as illustrated in the figure. 
Given the staggered configuration in the left panel of 
Fig.~\ref{fig: staggered-Ising}, and imposing periodic boundary conditions, 
there are $L/2$ such diagonals 
(recall that $L$ must be even to ensure bipartiteness). 
For any such diagonal line, we are free to take the dimers that intersect 
the diagonal and rotate them by $\pi/2$ 
without violating the dimer constraint or creating flippable plaquettes, 
as it has been done on the right panel of Fig.~\ref{fig: staggered-Ising}. 
Hence, both configurations depicted in Fig.~\ref{fig: staggered-Ising}
are degenerate in energy when $J/T=0$. 
By global $\pi/2$ rotations of all the
bonds intersecting any one of the $L/2$ diagonals, one generates a total
of $2^{L/2}$ degenerate configurations when $J/T=0$~\cite{Batista2004}. 
This degeneracy equals the number of all possible Ising spin configurations 
along a chain made of $L/2$ sites. 
Notice that we could have equivalently well chosen a top-left to bottom-right 
diagonal instead of a top-right to bottom-left diagonal as illustrated 
in Fig.~\ref{fig: staggered-Ising}. This choice can be shown to generate 
another $2^{L/2}$ degenerate configurations, only one of which has already 
been counted before (for a total of $2 \times 2^{L/2}-1$). 
Similarly, one can show that if we take as the starting configuration the 
one that obtains from the left panel of Fig.~\ref{fig: staggered-Ising} by 
means of a translation by one lattice spacing in the horizontal (or vertical) 
direction, the same procedure would yield another $2 \times 2^{L/2}-3$ 
independent degenerate configurations 
(for a grand total of $4 \times 2^{L/2}-4$). 

Switching on the interaction $J$ in
Eq.~(\ref{eq: J interaction}) 
lifts this macroscopic degeneracy by assigning an energy 
\begin{equation}
E^{(J)}_{\mathcal{C}_{\textrm{stagg}}} 
= 
E^{(J,\parallel)}_{\mathcal{C}_{\textrm{stagg}}} 
+
E^{(J,\perp)}_{\mathcal{C}_{\textrm{stagg}}}
\end{equation}
to each configuration in this manifold. 
The energy $E^{(J)}_{\mathcal{C}}$ has been decomposed into 
two contributions.
The first one,
\begin{equation}
\begin{split}
E^{(J,\parallel)}_{\mathcal{C}_{\textrm{stagg}}}
&:= 
-\frac{3}{2}J L\times (L/2)
\\
&=
-\frac{3}{2}\sum_{i=1}^{L/2} J L
\end{split}
\end{equation}
does not lift the degeneracy of this manifold
and can be thought of as resulting from the zero-point energy
of a diagonal labeled by the integer $i=1,\cdots,L/2$.
The second one, 
\begin{equation}
\begin{split}
E^{(J,\perp)}_{\mathcal{C}_{\textrm{stagg}}} 
&:=
-J L \sum^{L/2}_{i=1} 
  \left( 
    \frac{1 + S^{\ }_i S^{\ }_{i+1}}{2}
  \right)
+
\frac{J L}{2}\times(L/2) 
\\
&=
-
\frac{JL}{2} 
\sum^{L/2}_{i=1} S^{\ }_i S^{\ }_{i+1}, 
\end{split}
\label{eq: def 1D Ising}
\end{equation}
can be thought of as resulting from an Ising interaction between
consecutive diagonals whereby the Ising spin $S^{\ }_{i}$ 
variable takes the value $+1$ $(-1)$
if the diagonal intersects vertical (horizontal) dimers.
We have thus mapped the problem of 
evaluating the energy~(\ref{eq: J interaction})
for all configurations obtained from the seed configuration in 
Fig.~\ref{fig: staggered-Ising}
by flipping rigidly all the dimers that intersect 
top-right to bottom-left diagonals as
in Fig.~\ref{fig: staggered-Ising}
into an effective one-dimensional Ising model with the extensive
exchange energy $JL/2$.
Observe that the fact that the nearest-neighbor Ising interaction is 
proportional to the system size in the $1D$ Ising 
model~(\ref{eq: def 1D Ising}) implies that (a) the 
system orders for any non-vanishing value of the dimensionless ratio $J/T$
and
(b) excitations are separated from the ordered state by an energy gap
that grows linearly with $L$. 
The Ising order is ferromagnetic when $J>0$ 
and antiferromagnetic when $J<0$. 
The dimers are all aligned along the same direction when $J/T>0$ 
(parallel staggered order). 
When $J/T<0$, the dimers are aligned along the same direction within a 
diagonal but are rotated by $\pi/2$ between any two consecutive diagonals 
(alternating staggered order). 
Equivalent results are obtained if we consider top-left to bottom-right 
diagonals or a translated starting onfiguration, as discussed above. 
Therefore, the ground state manifold of energy~(\ref{eq: J interaction}) 
is four-fold degenerate for any finite $J$. 


Next, we consider the case of $u/T\gg1$ and $J/T=0$
when the system lies deep in the columnar phase.
There are four degenerate columnar ground states
one of which is depicted in the left panel of Fig.~\ref{fig: columnar-Ising}
and the other three are obtained upon rotating globally all the dimers by 
$\pi/2$ or upon a global translation by one lattice spacing. 
This degeneracy is preserved when switching on the interaction $J$, which 
results in the energy
\begin{equation}
E^{(J)}_{\mathcal{C}_{\textrm{col}}} 
= 
-\frac{J L^2}{2} 
\geq 
E^{(J)}_{\mathcal{C}_{\textrm{stagg}}}, 
\qquad\qquad
\left( \textrm{for } \; J>0 \right).
\label{eq: col J energy}
\end{equation}
The equality holds only for the four staggered configurations with 
alternating dimer directions along any two consecutive diagonals. 

\begin{figure}[ht]
\centering
\includegraphics[width=0.6\columnwidth]{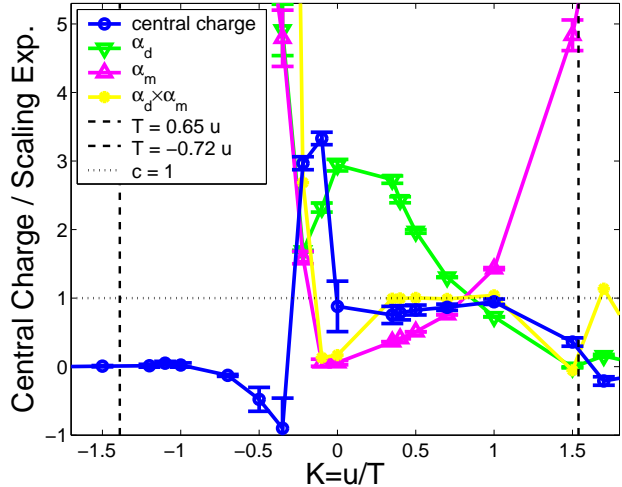}
\caption{
\label{fig: c d uJ05}
Central charge and scaling exponents for the electric and magnetic 
monopole operators as a function of $K=u/T$. 
Positive values of $K$ correspond to the columnar side of the interaction, 
while negative ones correspond to the staggered side. 
Following the convention in Ref.~\onlinecite{Alet2005}, 
we plot the scaling exponents $\alpha^{\ }_d = 2\,d^{\ }_{1,0}$ and 
$\alpha^{\ }_m = 2\,d^{\ }_{0,1}$ (equal to $1/g$ and $g$ from 
Eq.~(\ref{eq: scaling dimensions}), respectively), 
instead of the scaling dimensions $d^{\ }_{e,m}$. 
The value of the reduced coupling constant $J/T = 0.5$ is held constant 
throughout. 
The dashed lines indicate the location of the staggered and columnar 
transitions in the $J=0$ case discussed in Sec.~\ref{sec: results}. 
}
\end{figure}

Third, we want to estimate the average energy 
$E^{(J)}_{{\mathrm{dis}}}$ of configurations in the 
disordered phase due to the coupling $J$. 
This is given by 
\begin{equation}
E^{(J)}_{{\textrm{dis}}}
= 
-\frac{J}{2}
\left(
2\times2\times
\mathcal{C}^{\ }_{\textrm{nn}}
+
2\times4\times
\mathcal{C}^{\ }_{\textrm{nnn}}
\right)
L^{2}, 
\end{equation}
where $\mathcal{C}^{\ }_{\textrm{nn}}$ is the probability to find
a pair of parallel nearest-neighbor dimers and
$\mathcal{C}^{\ }_{\textrm{nnn}}$ is the probability to find
a pair of parallel next-nearest-neighbor dimers 
in the disordered phase. 
In the non-interacting limit $u,J \to 0$, the probabilities 
$\mathcal{C}^{\ }_{\textrm{nn}}$ and $\mathcal{C}^{\ }_{\textrm{nnn}}$
can be obtained exactly in the thermodynamic limit, as illustrated 
by M.~E.~Fisher and J.~Stephenson in Ref.~\onlinecite{Fisher1961} 
\begin{equation}
\mathcal{C}^{\ }_{\textrm{nn}} 
= 
\frac{1}{8}, 
\qquad\qquad
\mathcal{C}^{\ }_{\textrm{nnn}} 
= 
\frac{1}{4}
\left(
  \frac{1}{2} 
  - 
  \frac{1}{\pi} 
\right).
\end{equation}
Consequently, to leading order in $L$, one obtains 
\begin{equation}
E^{(J)}_{{\textrm{dis}}} 
= 
-
J 
\left[
  \frac{2L^2}{8} 
  + 
  \frac{4L^2}{4} 
    \left( 
      \frac{1}{2} 
      - 
      \frac{1}{\pi} 
    \right)
\right]
= 
-\frac{JL^2}{2} \frac{(3\pi - 4)}{2\pi} 
\end{equation}
in the limit $u,J \to 0$. 
Notice that if $J>0$ then 
$
E^{(J)}_{{\textrm{dis}}} 
> 
E^{(J)}_{\mathcal{C}_{\textrm{col}}} 
\geq 
E^{(J)}_{\mathcal{C}_{\textrm{stagg}}} 
$. 
For finite values of the couplings $J$ and $u$, the probabilities 
$\mathcal{C}^{\ }_{\textrm{nn}}$ and $\mathcal{C}^{\ }_{\textrm{nnn}}$
in the disordered phase are modified and the above result no longer holds 
exactly. However, we conjecture that the inequality 
$
E^{(J)}_{{\textrm{dis}}} 
> 
E^{(J)}_{\mathcal{C}_{\textrm{stagg}}}
$ 
holds true throughout the line of critical points separating the columnar 
phase from the staggered phase for $J>0$. 
This conjecture is indeed confirmed by our numerical transfer matrix 
results (e.g., see Fig.~\ref{fig: c d uJ05}) that clearly indicate a 
shrinking of the disordered phase for positive values of $J$. 

For sufficiently large values of $J/T$, 
the disordered phase is expected to disappear as $J$ favors 
the parallel alignment of nearest-neighbor and next-nearest-neighbor 
dimers. 
The transition between the staggered and columnar phases is then 
controlled by the competition between 
the dimensionless free energies of the two ordered phases 
\begin{equation}
\label{eq: free en crossing}
\begin{split}
F_{\textrm{col}}(T) 
&= 
-\frac{u L^2}{2T} -\frac{J L^2}{2T} - \ln(4),
\\ 
F_{\textrm{stagg}}(T)
&= 
-\frac{J L^2}{T} - \ln(4), 
\end{split}
\end{equation}
where for the staggered case we used the energy and entropy consistent 
with the $J>0$ case. 
This yields an estimate for the transition to occur when $J = u$ 
that is in remarkably good agreement with the numerical results. 

Let us now turn to the TM calculations. 
When $J>0$, the $W^{\ }_x=0$ sector captures the physics of the global 
transfer matrix much like in the $J=0$ case and we can thus use the 
same type of approach. 
The critical nature of the disordered phase appears to be preserved by the 
$J$ coupling, although its width is gradually reduced in agreement with our 
conjectured phase diagram in Fig.~\ref{fig: uJ phase diag}. 
For brevity, we show in Fig.~\ref{fig: c d uJ05} 
only the central charge and scaling exponents obtained for $J/T=0.5$.
The origin of the peak in the central charge appearing at the first order 
transition to the staggered phase is as of now unclear, though it is likely 
due to some spurious finite size effect (see also the discussion 
below on the first order transition between the columnar and staggered 
phase at low temperatures). 
For values of the reduced coupling $J/T$ larger than 
$(J/T)^{\ }_c \simeq 0.6$ the critical region shrinks to the point 
$J=u$ at which a direct transition between the staggered and the 
columnar phase takes place. 
This is in agreement with the free energy arguments 
of Eq.~(\ref{eq: free en crossing}), suggesting that the transition 
arising from the collapse of the line of critical points is indeed a 
first order transition. 
Observe that the first-order critical point $J=u$ that appears at low 
temperatures ($J/T \gtrsim 0.6$) is special from an entropic point of view. 
Indeed, all the configurations 
obtained from a columnar one through any combination of the translations 
illustrated in Fig.~\ref{fig: columnar-Ising} are degenerate in energy 
when $J=u$. 
While the number of degenerate lowest-energy configurations is finite in 
the parallel staggered and columnar phase, it becomes exponential in the 
linear size of the system at the $J=u$ critical point. 
Namely, there are $2^L+2^M$ degenerate configurations in an $L \times M$ 
system, whereby $2^L$ ($2^M$) correspond to having all the dimers 
aligned horizontally (vertically). 
Consequently, at the coexistence point $J=u$ the intensive entropy of 
the system scales with system size as 
\begin{equation}
\frac{\ln \left( e^L + e^M \right)}{L M}.
\end{equation}
In turn,
this gives a contribution to the dimensionless intensive free energy 
obtained from TM calculations of the form 
\begin{equation}
\lim_{M\to\infty} \frac{\ln \left( e^L + e^M \right)}{L M} 
= 
\frac{1}{L}. 
\label{eq: 1over L correction}
\end{equation}
Equation~(\ref{eq: 1over L correction})
introduces a $1/L$ finite-size scaling correction, 
the signature of which is to give rise to a peak in the central charge at 
$J=u$
($c$ would actually be infinite if the $L\to\infty$ extrapolation could 
be done exactly) when using the $1/L^{2}$
finite size scaling Ansatz for the central charge 
as in Sec.~\ref{sec: results}.
We can thus take advantage of this effect, 
when fitting the free energy as in 
Eqs.~(\ref{eq: scaling forms}) and~(\ref{eq: pow law}), 
to interpret the peak in the free energy when $J/T\gtrsim 0.6$
as a signature of a first order phase transition.
A remnant of the $1/L$ correction to scaling might also be responsible for the 
central charge peak at the staggered transition 
that appears in the numerical results when $J/T\lesssim 0.6$
(see Fig.~\ref{fig: c d uJ05}). 

\begin{figure}[ht]
\centering
\includegraphics[width=0.6\columnwidth]{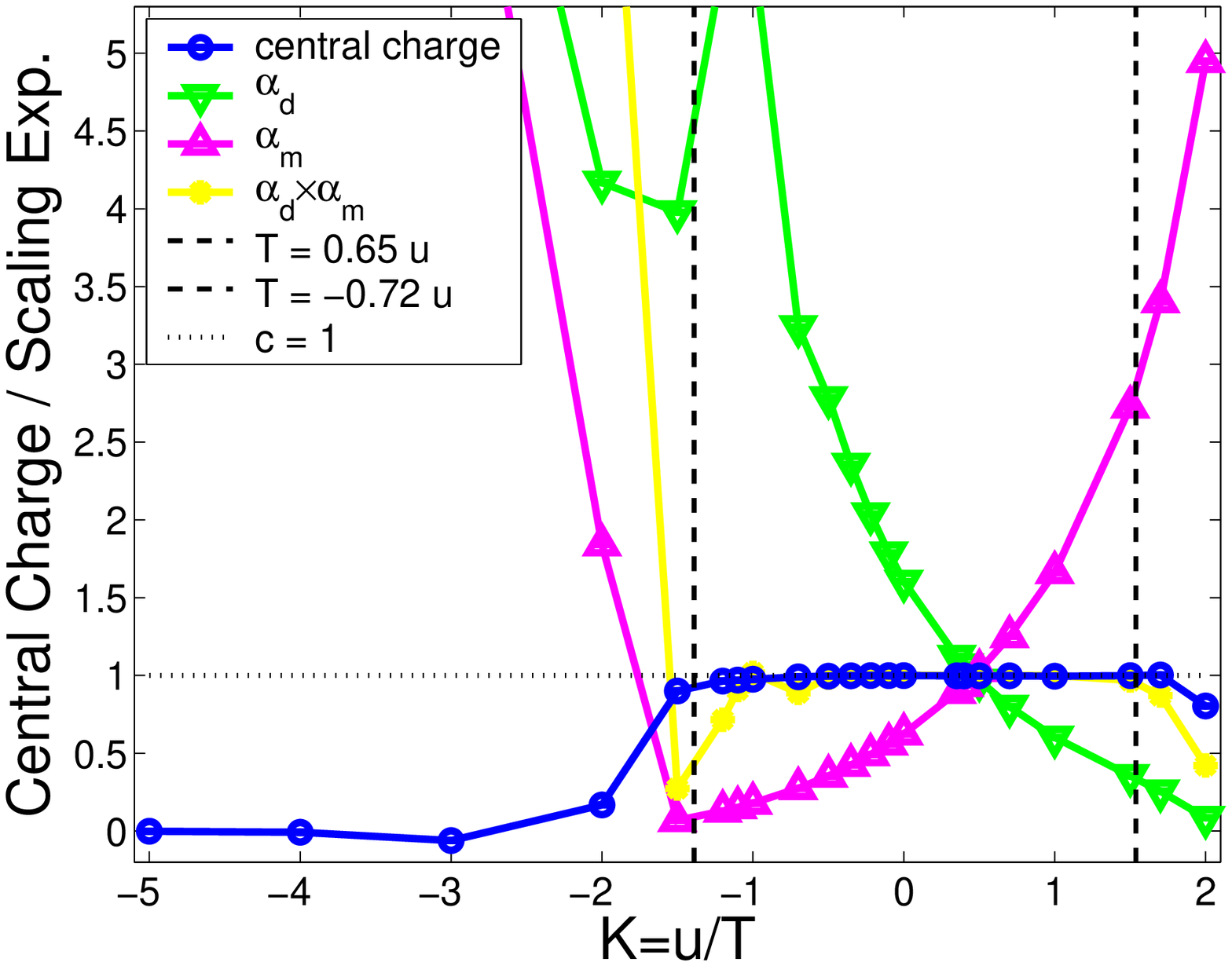}
\includegraphics[width=0.6\columnwidth]{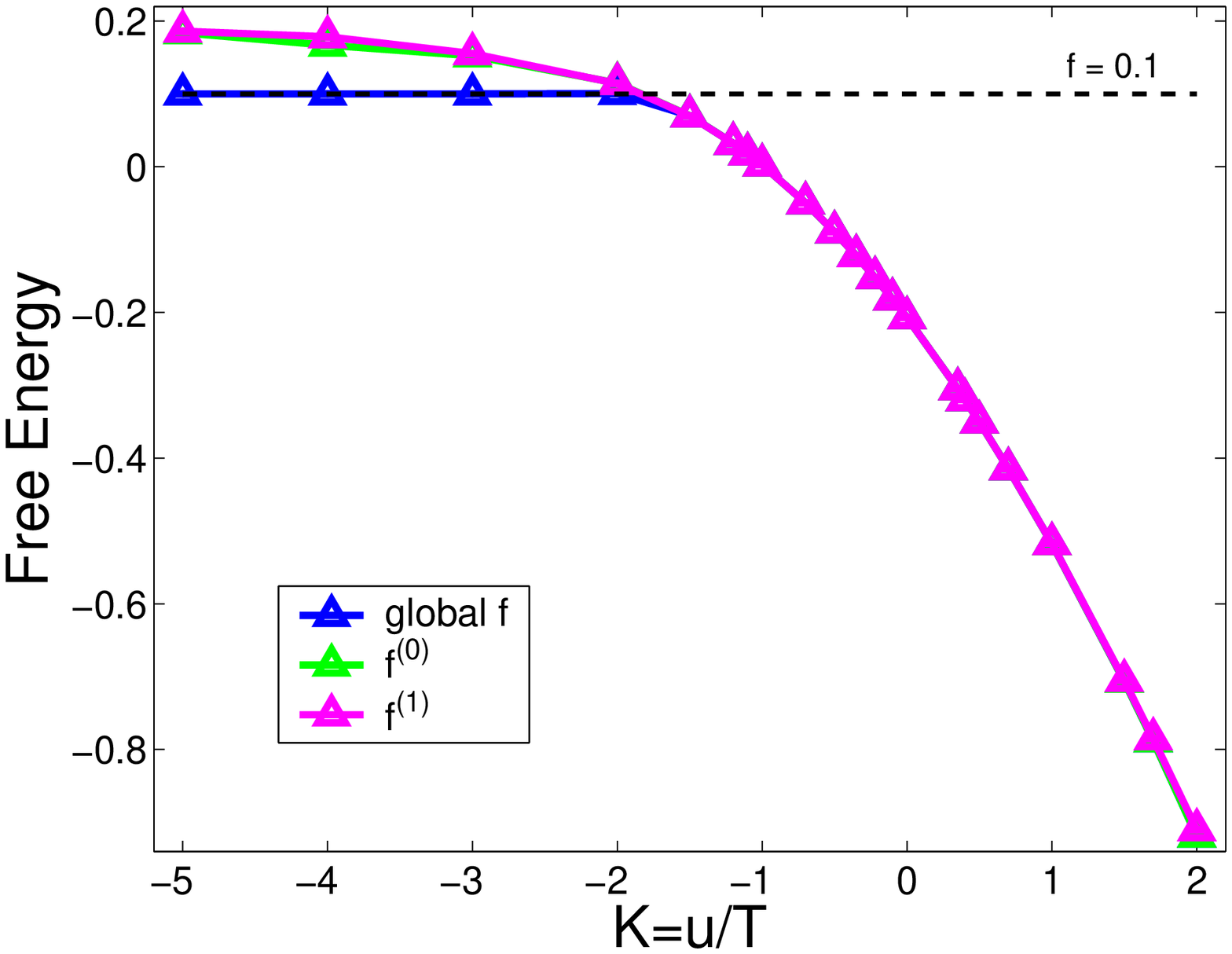}
\caption{
\label{fig: c d uJ_02dg}
(Top) 
Central charge and scaling exponents for the electric and magnetic 
monopole operators as a function of $K=u/T$. 
Positive values of $K$ correspond to the columnar side of the interaction, 
while negative ones correspond to the staggered side. 
Following the convention in Ref.~\onlinecite{Alet2005}, 
we plot the scaling exponents $\alpha^{\ }_d = 2\,d^{\ }_{1,0}$ and 
$\alpha^{\ }_m = 2\,d^{\ }_{0,1}$ (equal to $1/g$ and $g$ from 
Eq.~(\ref{eq: scaling dimensions}), respectively), 
instead of the scaling dimensions $d^{\ }_{e,m}$. 
The value of the reduced coupling constant $J/T = -0.2$ is held constant 
throughout. 
The dashed lines indicate the location of the staggered and columnar 
transitions in the $J=0$ case discussed in Sec.~\ref{sec: results}. 
(Bottom) 
Free energy of the system obtained from the global TM and from the 
TM restricted to the $W^{\ }_x=0$ and $W^{\ }_x=1$ sectors respectively. 
Notice that only the global free energy saturates at $-J/2=0.1$ 
for large negative values of $u/T$, as expected from 
Eq.(~\ref{eq: col J energy}). 
}
\end{figure}

When $J<0$, the numerical analysis becomes more delicate because the 
staggered order that the system develops at large, negative 
values of $u/T$ has alternating dimer directions between consecutive 
diagonals [see Fig~\ref{fig: staggered-Ising}]. 
Studying this phase demands larger lattices due to the symmetry breaking 
pattern which allows only values of $L$ that 
are multiple of $4$ in the notation used in Fig.~\ref{fig: TM labeling}. 
Moreover, this type of order lies in the $W^{\ }_{x}=\pm L/4$ sector and 
the transition cannot be captured by considering the 
$W^{\ }_x=0$ sector alone (which is the correct one for the 
columnar transition). 
In Fig.~\ref{fig: c d uJ_02dg} we present the 
results obtained from the $L=8,12,16$ data with the global TM. 
The small number of accessible system sizes does not allow for an 
appropriate $L\to\infty$ extrapolation~(\ref{eq: pow law}) and the values 
in Fig.~\ref{fig: c d uJ_02dg} are directly obtained from the finite size 
scaling fits~(\ref{eq: scaling forms}). 
As expected, upon changing the sign of the coupling $J$ the disordered 
phase is now favored over the two ordered ones and the line of critical 
points appears to expand with respect to the $J=0$ case in 
Fig.~\ref{fig: c d uJ_02dg} (Top). 
{F}rom the behavior of the global and restricted free energies one sees
that the correct asymptotic behavior~(\ref{eq: col J energy}) 
$\lim_{u/T\to-\infty}(f)=-J/2=0.1$ is captured only by the global TM. 
%
%

\section{\label{sec: monomers}
The role of defects
        }
The purpose of this section is to generalize Hamiltonian~(\ref{eq: def H0})
in such a way that the dynamics of defects becomes possible without spoiling
the existence of a GS of the form~(\ref{eq: quantum Alet GS}).
We shall only be concerned with point defects that we call monomers.
A monomer is a site of the square lattice that is not the end point
of a dimer. 

\begin{figure}[t]
\centering
\includegraphics[width=0.4\columnwidth]{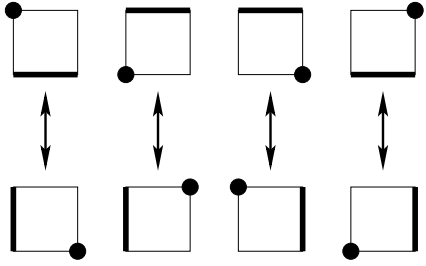}
\caption{
\label{fig: eight is 4 pair of moves}
There are eight ways for a plaquette of the square lattice to accommodate
a single monomer (a site occupied by a filled circle) and a single dimer 
on any of its two remaining free edges. These eight configurations
are grouped into pairs
that define the local moves in 
configuration space. Any of these local moves amounts to a 
rotation by $\pi/2$ of the dimer and a diagonal hop of the monomer. 
        }
\end{figure}

\begin{figure}[t]
\centering
\includegraphics[width=0.74\columnwidth]{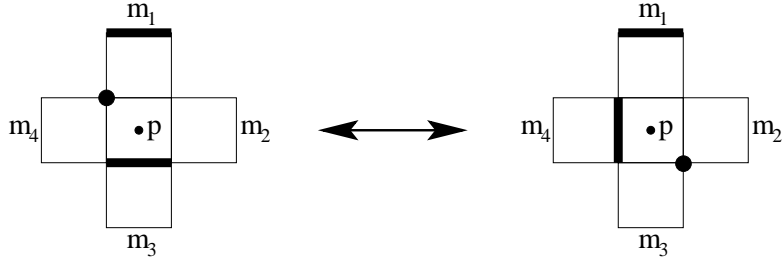}
\caption{
\label{fig: decor_plaq_mon}
One of the two pairs of decorated plaquettes 
$\ell^{*}_{1}$
and
$\overline{\ell^{*}_{1}}$ 
that involve hopping of the monomer between the upper left and lower 
right corners of the plaquette $p$.
        }
\end{figure}

\begin{figure}[t]
\centering
\includegraphics[width=0.55\columnwidth]{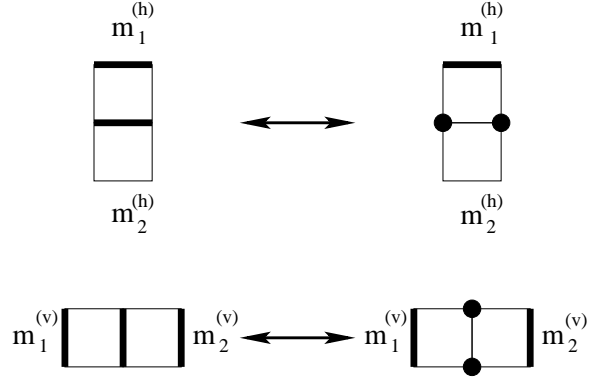}
\caption{
\label{fig: decor_bonds}
A pair of decorated horizontal bonds 
$\ell^{*(h)}_{2}$ and $\overline{\ell^{*(h)}_{2}}$
and a pair of decorated vertical bonds
$\ell^{*(v)}_{2}$ and $\overline{\ell^{*(v)}_{2}}$
that are related by the process of
creation~/~annihilation of a pair of monomers.
The information encoded by the decorated bond is the location of the bond,
whether it is occupied by a dimer or two monomers, and whether the two
bonds facing it are occupied by a dimer ($m^{(h,v)}_{1,2}=1$)
or not ($m^{(h,v)}_{1,2}=0$). 
        }
\end{figure}

The strategy that we will follow consists of three steps.
First, we define the enlarged classical configuration space 
$\mathcal{S}$ describing a \emph{dilute} SLDM where each site belongs to 
\emph{at most} one dimer. 
Second, we define the classical configuration energy
\begin{equation}
E^{(u,\mu)}_{\mathcal{C}} := 
-
\left(
u\,N^{(f)}_{\mathcal{C}} 
+ 
\mu\,M^{\ }_{\mathcal{C}}
\right)
\end{equation}
where $N^{(f)}_{\mathcal{C}}$ is, as before, the total number of plaquettes 
occupied by two parallel dimers in configuration $\mathcal{C}$ while 
$M^{\ }_{\mathcal{C}}$ is the total number of monomers in configuration 
$\mathcal{C}$.
The energy scale $\mu\in\mathbb{R}$ 
thus plays the role of a chemical potential for the monomers.
Third, we need to generalize the local moves 
$\ell^{*}_{0} \leftrightarrow \overline{\ell^{*}_{0}}$ 
in the classical configuration space $\mathcal{S}^{\ }_{0}$
that we encountered in Fig.~\ref{fig: decorated plaquette}
to endow monomers with quantum dynamics.
To this end, we first define \textit{decorated} plaquettes 
that involve monomers.
Any plaquette of the square lattice has four ways to accommodate a single
monomer. Given a plaquette occupied by a single monomer, there are two ways
to accommodate a dimer on one of its two remaining free edges.
Any plaquette of the square lattice has thus eight ways 
to accommodate a single monomer and a single dimer. 
In turn, these eight ways can be grouped into pairs 
each of which defines a local move 
in the classical configuration space as 
shown in Fig.~\ref{fig: eight is 4 pair of moves}.
A decorated plaquette 
$\ell^{*}_{1}$ 
at $p$ built around a single 
monomer and a single dimer 
is obtained by taking Fig.~\ref{fig: decorated plaquette}
and replacing the flippable plaquette $p$ by any of the
plaquettes in Fig.~\ref{fig: eight is 4 pair of moves}
as is done in Fig.~\ref{fig: decor_plaq_mon}.
Then, we want to account for the creation and annihilation of pairs
of monomers. Following the same steps as above, we introduce the 
decorated horizontal and vertical bonds 
$\ell^{*(h)}_{2}$ 
and 
$\ell^{*(v)}_{2}$, 
as well as the corresponding local updates 
$
\ell^{*(h)}_{2}
\leftrightarrow
\overline{\ell^{*(h)}_{2}}
$
and
$
\ell^{*(v)}_{2}
\leftrightarrow
\overline{\ell^{*(v)}_{2}}
$
defined in Fig.~\ref{fig: decor_bonds}. 
For notational simplicity we will omit in the sequel the reference to the 
bond orientation $h,v$, as we have already done for the bond $b$ and 
plaquette $p$ position labels, and for the eight different types of decorated 
monomer-dimer plaquettes. 

The new quantum Hamiltonian can then be assembled from three separate parts, 
the pure dimer contribution $\widehat{H}^{\ }_{0}$ and two terms arising from 
the introduction of monomers, $\widehat{H}^{\ }_{1}$ and 
$\widehat{H}^{\ }_{2}$, 
\begin{eqnarray}
\widehat{H}^{\ }_{\textrm{\small tot}} 
&=& 
\widehat{H}^{\ }_{0} + \widehat{H}^{\ }_{1} + \widehat{H}^{\ }_{2} 
\nonumber\\ 
&=& 
\frac{t^{\ }_{0}}{2} \sum_{\ell^{*}_{0}} 
  \left[ \vphantom{\Big[}
    e^{u \delta N^{(f)}_{\ell^{*}_{0}}/2T} 
      \vert\ell^{*}_{0}\rangle\langle\ell^{*}_{0}\vert 
    + 
    e^{u \delta N^{(f)}_{\overline{\ell^{*}_{0}}}/2T} 
      \vert\overline{\ell^{*}_{0}}\rangle\langle\overline{\ell^{*}_{0}}\vert 
    - \left( \vphantom{\Big[}
        \vert\overline{\ell^{*}_{0}}\rangle\langle\ell^{*}_{0}\vert 
	+ 
        \vert\ell^{*}_{0}\rangle\langle\overline{\ell^{*}_{0}}\vert 
      \right)
  \right] 
\nonumber\\
&+& 
\frac{t^{\ }_1}{2} \sum_{\ell^{*}_{1}} 
  \left[ \vphantom{\Big[}
    e^{u \delta N^{(f)}_{\ell^{*}_{1}}/2T} 
      \vert\ell^{*}_{1}\rangle\langle\ell^{*}_{1}\vert 
    + 
    e^{u \delta N^{(f)}_{\overline{\ell^{*}_{1}}}/2T} 
      \vert\overline{\ell^{*}_{1}}\rangle\langle\overline{\ell^{*}_{1}}\vert 
    - \left( \vphantom{\Big[}
        \vert\overline{\ell^{*}_{1}}\rangle\langle\ell^{*}_{1}\vert 
	+ 
        \vert\ell^{*}_{1}\rangle\langle\overline{\ell^{*}_{1}}\vert 
      \right)
  \right] 
\nonumber\\
&+& 
\frac{t^{\ }_2}{2} \sum_{\ell^{*}_{2}} 
  \left[ \vphantom{\Big[}
    e^{(u \delta N^{(f)}_{\ell^{*}_{2}} + \mu 
       \delta M^{\ }_{\ell^{*}_{2}})/2T} 
      \vert\ell^{*}_{2}\rangle\langle\ell^{*}_{2}\vert 
    + 
    e^{(u \delta N^{(f)}_{\overline{\ell^{*}_{2}}} + 
        \mu \delta M^{\ }_{\overline{\ell^{*}_{2}}})/2T} 
      \vert\overline{\ell^{*}_{2}}\rangle\langle\overline{\ell^{*}_{2}}\vert 
\right.
\nonumber\\
&& 
\hphantom{
\qquad\qquad\qquad\qquad\qquad\qquad\qquad\qquad\quad\,
         }
\left.
    - \left( \vphantom{\Big[}
        \vert\overline{\ell^{*}_{2}}\rangle\langle\ell^{*}_{2}\vert 
	+ 
        \vert\ell^{*}_{2}\rangle\langle\overline{\ell^{*}_{2}}\vert 
      \right)
  \vphantom{\Big[} \right].
\nonumber\\
&&
\label{eq: quantum Alet + monomer Hamiltonian}
\end{eqnarray}
The energy scales $t^{\ }_0$, $t^{\ }_1$, and $t^{\ }_2$ are positive. 
For any $n\in\{0,1\}$, the integer
$
\delta N^{(f)}_{\ell^{*}_{n}}=
-
\delta N^{(f)}_{\overline{\ell^{*}_{n}}}
$
is nothing but the change
$
N^{(f)}_{\overline{\mathcal{C}}}
-
N^{(f)}_{\mathcal{C}}
$
in the number of flippable plaquettes induced by the
local update $\ell^{*}_{n}\leftrightarrow\overline{\ell^{*}_{n}}$ 
when the action of $\vert\ell^{*}_{n}\rangle\langle\ell^{*}_{n}\vert$ 
on the state $\vert\mathcal{C}\rangle$ is non-vanishing
(or, equivalently, when the action of
$\vert\overline{\ell^{*}_{n}}\rangle\langle\overline{\ell^{*}_{n}}\vert$ 
on the updated state $\vert\overline{\mathcal{C}}\rangle$ 
is non-vanishing). The integer
$
\delta M^{\ }_{\ell^{*}_{2}}=
-
\delta M^{\ }_{\overline{\ell^{*}_{2}}}
$ takes the value $+2$ when
two monomers are created and the value $-2$ when two monomers 
are annihilated under the local update 
$\ell^{*}_{2}\leftrightarrow\overline{\ell^{*}_{2}}$. 
We leave it as an exercise to the reader to express
$\delta N^{(f)}_{\ell^{*}_{n}}$ in terms of the integers
$m^{\ }_{1,2,3,4}$ defined in Fig.~\ref{fig: decor_plaq_mon}
and in terms of the integers
$m^{(h,v)}_{1,2}$ defined in Fig.~\ref{fig: decor_bonds}.

As desired, the GS of Hamiltonian%
~(\ref{eq: quantum Alet + monomer Hamiltonian})
is given by
\begin{equation}
\vert\Psi^{\ }_{\textrm{\small tot}}\rangle 
= 
\sum_{
\mathcal{C}\in
\mathcal{S}^{\ }_{} 
     } 
  e^{(u N^{(f)}_{\mathcal{C}} + \mu M^{\ }_{\mathcal{C}})/2T} 
    \vert\mathcal{C}\rangle
\label{eq: quantum Alet + monomer GS}
\end{equation}
where the summation has been extended to account for the enlarged Hilbert 
space. 
By construction monomers always occur in pairs with one half 
of the monomers residing on one
of the sublattice of the square lattice. 
The zero-temperature phase diagram
of the quantum Hamiltonian~(\ref{eq: quantum Alet + monomer Hamiltonian}) 
thus contains the finite-$T$ phase diagram of the classical system with the
partition function
\begin{equation}
Z(T/u,\mu/T) := 
\sum_{
\mathcal{C}\in
\mathcal{S}^{\ }_{}
     }
  \exp \left(-\frac{E^{(u,\mu)}_{\mathcal{C}}}{T} \right) 
= 
\sum_{
\mathcal{C}\in
\mathcal{S}^{\ }_{}
     }
  \exp 
    \left(
      \frac{u N^{(f)}_{\mathcal{C}} + \mu M^{\ }_{\mathcal{C}}}{T} 
    \right). 
\label{eq: Alet + monomer partition function}
\end{equation}

\begin{figure}[ht!]
\centering
\includegraphics[width=0.7\columnwidth]{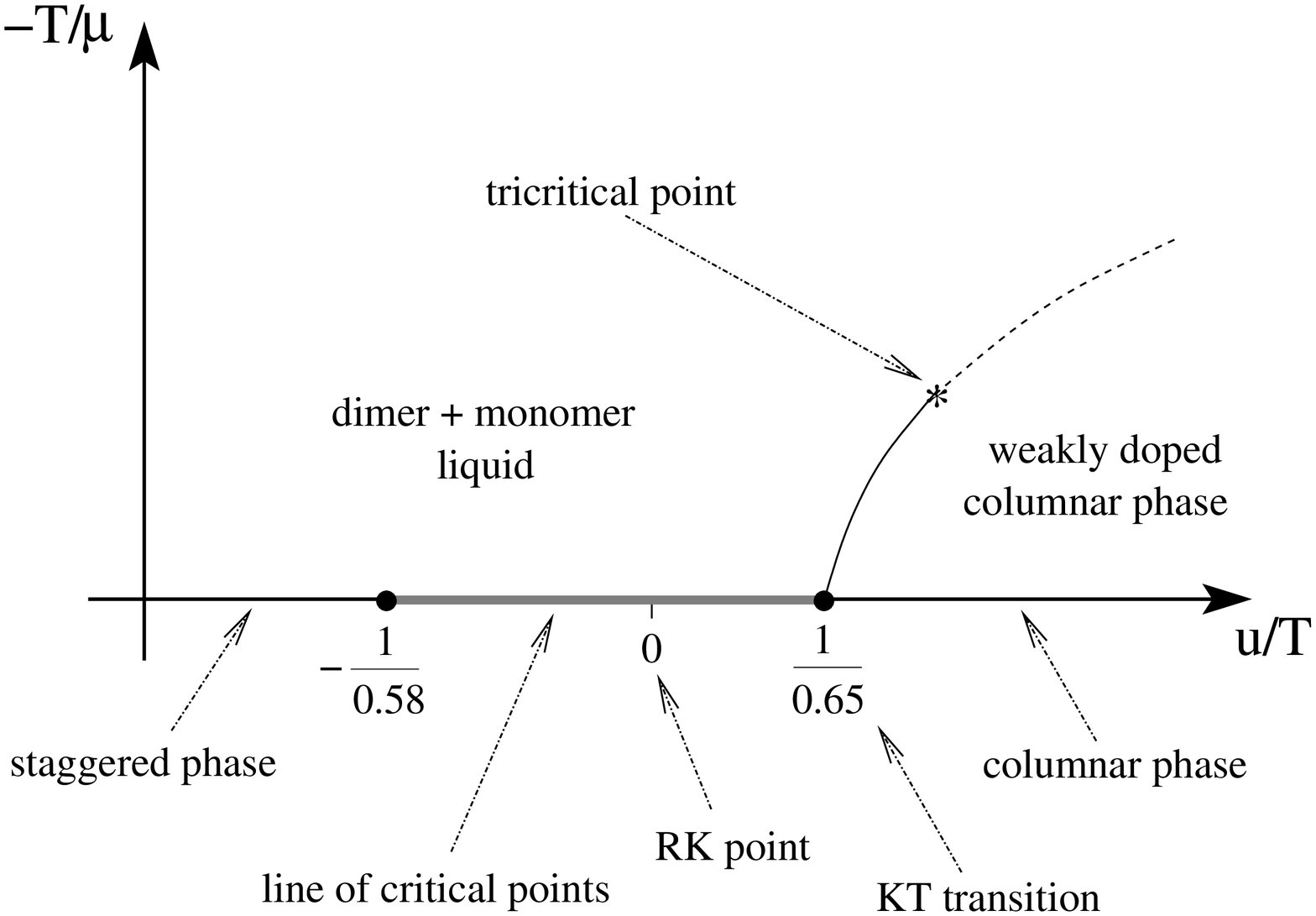}
\caption{
\label{fig: monomer phase diag}
Conjectured phase diagram of the classical partition 
function~(\ref{eq: Alet + monomer partition function}) after 
Alet~\textit{et al.} in Ref.~\onlinecite{Alet2005} that is
inherited by the SMF quantum dimer 
model~(\ref{eq: quantum Alet + monomer GS}). 
While for $u/T < u/T^{\textrm{\scriptsize (columnar)}}_c$ the 
introduction of monomers is a relevant perturbation -- 
which destroys the critical phase -- it becomes marginal at 
$T^{\textrm{\scriptsize (columnar)}}_c$. 
Therefore, a phase boundary $\mu(T)$ is expected to depart from the KT 
transition point at 
$(T,\mu)=(T^{\textrm{\scriptsize (columnar)}}_c,-\infty)$. 
In the close vicinity of 
$(T,\mu)=(T^{\textrm{\scriptsize (columnar)}}_c,-\infty)$, 
this phase boundary separates a weakly doped columnar ordered 
phase from a dilute (non-critical) dimer liquid 
and the transition between these two phases 
is expected to be continuous (solid line at the phase boundary). 
For both $-T/\mu$ and $u/T$ large, the phase boundary $\mu(T)$ separates 
phases of matter with large gaps and the transition between these two 
phases is expected to become first order (dashed line at the phase boundary). 
A \emph{tricritical point} at some finite 
$(T^{\ }_*,\mu^{\ }_*)$ must then separate the first order from the 
continuous behavior along the phase boundary. 
Alet~\textit{et al.} give an estimate for the tricritical point at 
$T^{\ }_*/u = 0.39(4)$  and 
$\mu^{\ }_*/u = 0.25(10)$~\cite{Alet2005}. 
        }
\end{figure}

A detailed study of the phase diagram of this partition function is beyond 
the scope of the present paper. 
However, according to some preliminary results by 
Alet~\textit{et al.} in Ref.~\onlinecite{Alet2005}, 
we anticipate a rich and interesting structure, qualitatively illustrated in 
Fig.~\ref{fig: monomer phase diag}. 
Notice that the $u=0$, $\mu=-\infty$ limit realizes precisely the 
RK point~(\ref{eq: def RK point}). 
Also, although the effect of a finite concentration of static monomers on 
the classical critical phase is an interesting open problem, one can 
reasonably expect that the spatial dimer correlations will be preserved below 
a cutoff length scale dictated by the average monomer-monomer separation. 
Therefore, the limit $t^{\ }_2=0$ realizes a classical neutral gas of 
charged hard-core particles (the monomers) at a fixed density, 
coupled to a \emph{quasi-critical} environment (the dimers). 

Finally, having constructed a quantum dilute SLDM Hamiltonian using the 
Stochastic Matrix Form 
decomposition~(\ref{eq: quantum Alet + monomer Hamiltonian}) offers the 
advantage of obtaining (static) GS correlation functions directly from the
associated classical system, as pointed out in Sec.~\ref{sec: intro}. 
The possibility to use classical numerical techniques, such as Monte Carlo 
simulations and transfer matrix calculations, in the 
\emph{same number of dimensions} gives access to much larger system 
sizes than quantum techniques, such as quantum Monte Carlo or 
exact diagonalization routines, do. 
These classical results allow to contrast quantum dynamical correlation 
functions of monomers against static correlation functions inherited 
from the classical partition function%
~(\ref{eq: Alet + monomer partition function}). 
In this context, related quantum dimer models with mobile holes 
have been studied by exact-diagonalization, Green-function, and classical 
Monte Carlo techniques by Poilblanc and collaborators~\cite{Poilblanc2006}. 
%
%

\subsection{\label{sec: pure monomer Hamiltonian}
The non-interacting monomer gas in a critical background
           }
In closing, we would like to discuss briefly a special case of the 
Hamiltonian in Eq.~(\ref{eq: quantum Alet + monomer Hamiltonian}), 
namely the one we obtain upon choosing $t^{\ }_0 = t^{\ }_1 = 0$ and $u=0$. 
Given this choice, $t^{\ }_2$ becomes just a scale factor and can be 
conveniently set to $1$. Also, the information about the integers 
$m^{(h,v)}_{1,2}$ defined in Fig.~\ref{fig: decor_bonds} is no longer needed 
and can be dropped. 
The new Hamiltonian thus obtained can be written in a symbolic 
notation along the line of Eq.~(\ref{eq: RK Hamiltonian}) as 
\begin{equation}
\widehat{H}^{\ }_{\textrm{mon}} 
= 
\sum^{\ }_{b} 
  \left[ \vphantom{\Big[} 
    e^{\mu/T} 
    \vert
\setlength{\unitlength}{0.06mm}
\begin{picture}(90,51)(0,8) 
  \linethickness{0.5mm}
  \put(18,25){\line(1,0){54}} 
  \put(18,25){\circle*{20}}
  \put(72,25){\circle*{20}}
\end{picture}
\rangle \langle
\setlength{\unitlength}{0.06mm}
\begin{picture}(90,51)(0,8) 
  \linethickness{0.5mm}
  \put(18,25){\line(1,0){54}} 
  \put(18,25){\circle*{20}}
  \put(72,25){\circle*{20}}
\end{picture}
\vert_b
  + 
    e^{-\mu/T} 
    \vert
\setlength{\unitlength}{0.06mm}
\begin{picture}(90,51)(0,8) 
  \put(18,25){\circle*{20}}
  \put(72,25){\circle*{20}}
\end{picture}
\rangle \langle
\setlength{\unitlength}{0.06mm}
\begin{picture}(90,51)(0,8) 
  \put(18,25){\circle*{20}}
  \put(72,25){\circle*{20}}
\end{picture}
\vert_b
  - \left( \vphantom{\Big[} 
      \vert
\setlength{\unitlength}{0.06mm}
\begin{picture}(90,51)(0,8) 
  \put(18,25){\circle*{20}}
  \put(72,25){\circle*{20}}
\end{picture}
\rangle \langle
\setlength{\unitlength}{0.06mm}
\begin{picture}(90,51)(0,8) 
  \linethickness{0.5mm}
  \put(18,25){\line(1,0){54}} 
  \put(18,25){\circle*{20}}
  \put(72,25){\circle*{20}}
\end{picture}
\vert_b
  + 
      \vert
\setlength{\unitlength}{0.06mm}
\begin{picture}(90,51)(0,8) 
  \linethickness{0.5mm}
  \put(18,25){\line(1,0){54}} 
  \put(18,25){\circle*{20}}
  \put(72,25){\circle*{20}}
\end{picture}
\rangle \langle
\setlength{\unitlength}{0.06mm}
\begin{picture}(90,51)(0,8) 
  \put(18,25){\circle*{20}}
  \put(72,25){\circle*{20}}
\end{picture}
\vert_b
    \right) 
  \right], 
\label{eq: monomer gas Hamiltonian}
\end{equation}
where the summation is over all vertical and horizontal bonds $b$ and the 
Hilbert space is the same as for 
Eq.~(\ref{eq: quantum Alet + monomer Hamiltonian}). 
The exact GS of this Hamiltonian 
\begin{equation}
\vert\Psi^{\ }_{\textrm{mon}}\rangle 
= 
\sum_{
\mathcal{C}\in
\mathcal{S}^{\ }_{} 
     } 
  e^{\mu M^{\ }_{\mathcal{C}}/2T} 
    \vert\mathcal{C}\rangle
\label{eq: monomer gas GS}
\end{equation}
is then captured by the classical dilute dimer model in the sole presence 
of a chemical potential $\mu$ for the monomers, 
\begin{equation}
Z^{\ }_{\textrm{mon}} 
= 
\sum_{
\mathcal{C}\in
\mathcal{S}^{\ }_{} 
     } 
  e^{\mu M^{\ }_{\mathcal{C}}/T}.
\label{eq: monomer gas partition function}
\end{equation}

In the limit of $\mu/T\to -\infty$, the monomers behave like deconfined 
virtual particles in a critically correlated dimer background, and the GS 
wavefunction becomes exponentially close to the GS wavefunction of the 
RK Hamiltonian~(\ref{eq: RK Hamiltonian}) at the so called RK 
point~(\ref{eq: RK GS}). 
%
%

\section{\label{sec: conclusions}
Conclusions
        }
In this paper, we constructed an interacting quantum dimer model
with a Hamiltonian obeying a Stochastic Matrix Form decomposition%
~\cite{Castelnovo2005}.
This allowed us to define an associated classical system whose thermodynamic 
equilibrium correlation functions reflect the behavior of some equal-time 
ground state expectation values. 
Using this correspondence, we showed that the quantum system exhibits 
a line of quantum critical points that separates two ordered GS of the 
valence bond crystal type. 
One end point of this line of critical points
corresponds to KT transition in the associated classical system.
The other one is a first order phase transition.
The locality of our SMF quantum Hamiltonian can be used to argue that 
the dynamical exponent $z$ is finite, and likely to be $z=2$,
along the line of critical points. If so, one can deduce the power-law decay 
of quantum correlations in time from the behavior of the equal-time spatial 
correlations.
In this sense, the quantum system inherits a KT transition from its
associated classical system.

We also studied the robustness of this line of critical points to the 
introduction of longer-range -- yet finite -- dimer-dimer interactions. 
In particular we showed how, for some choice of the additional interaction, 
one can tune the width of the line of critical points until it gradually 
collapses onto a single, tricritical point, where the KT and first-order 
transition meet. 

Eventually, we included monomers in the quantum system, i.e., we allowed 
for sites that are not occupied by a dimer. In doing so, we preserved the 
SMF structure of the Hamiltonian, thus being able to study the GS 
properties of the quantum system via an associated classical dilute dimer 
model. Using this correspondence, we derived qualitatively the 
zero-temperature 
phase diagram of the quantum dilute dimer model, opening the possibility 
to obtain equal-time monomer correlation functions from classical 
numerical techniques. While this allows access to much larger system sizes 
to study zero-temperature spatial correlation functions, unequal-time monomer 
correlations remain the prerogative of quantum MC or exact 
diagonalization studies.
%
%

P.P.\ would like to thank Didier Poilblanc and Fabien Alet
for early discussions and sharing their results prior to publication.
C.C.\ would like to thank Cristian D.~Batista for explaining 
useful details of Ref.~\cite{Batista2004}. 
We would also like to thank Eduardo Fradkin for his insightful comments. 
%
%


\begin{thebibliography}{99}
\bibitem{note: staggered caveat}
The reasons for this choice of values of $L$ is similar to the 
one discussed in Sec.~\ref{sec: uJ results summary}. 
Namely, using the global TM and the full range of 
values of $L=6,8,10,12,14,16$ we noticed an oscillatory behavior of the 
estimants for the central charge $c(L^{\ }_{0},L^{\ }_{\textrm{\small max}})$ 
and scaling dimensions $d^{\ }_{e,m}(L^{\ }_{0},L^{\ }_{\textrm{\small max}})$ 
as a function of $L^{\ }_{0}$ (see Sec.~\ref{sec: results}). 
This usually signals the presence in the system of fluctuations 
with a spatial periodicity larger than the one generically allowed by $L$ 
(two lattice spacings). 
Choosing $L$ to be a multiple of $4$ seems to be sufficient in our case 
to get rid of this spurious behavior.
\bibitem{note: RK resonating}
Notice however that the two ordered phases separated by the tricritical 
point may differ from those in the RK Hamiltonian, if the presence of a 
resonating valence bond phase was to be confirmed in the latter model.
\end{thebibliography}
\end{document}